\documentclass[journal,10pt]{IEEEtran}
\IEEEoverridecommandlockouts

\usepackage{hyperref}
\usepackage{graphicx}
\usepackage{cite}
\usepackage{subfigure}
\usepackage{amsmath,amssymb,amsfonts}
\usepackage{bm}
\usepackage{booktabs}
\usepackage{multirow}
\usepackage{array}
\usepackage{stfloats}
\usepackage{xcolor}
\usepackage{url}
\usepackage{algorithmic}
\usepackage[ruled]{algorithm2e}
\usepackage[skip=1pt]{caption}
\newcommand{\bF}{\mathbf F}
\newcommand{\bI}{\mathbf I}

\newcommand{\bR}{\mathbf R}

\newtheorem{remark}{Remark}
\def\BibTeX{{\rm B\kern-.05em{\sc i\kern-.025em b}\kern-.08em
    T\kern-.1667em\lower.7ex\hbox{E}\kern-.125emX}}

\begin{document}

\title{SIM-Enabled Wave-Domain Processing in MIMO-OFDM Receivers: Architecture, Modeling, and Validation}
\author{Zheao Li, \IEEEmembership{Graduate Student Member, IEEE}, Jiancheng An, \IEEEmembership{Senior Member, IEEE}, and Chau Yuen, \IEEEmembership{Fellow, IEEE}
\vspace{-1 cm}
\thanks{Z. Li, J. An, and C. Yuen are with the School of Electrical and Electronic Engineering, Nanyang Technological University, 639798, Singapore (email: zheao001@e.ntu.edu.sg, jiancheng.an@uestc.edu.cn, chau.yuen@ntu.edu.sg). (\emph{Corresponding author: Chau Yuen}.)}
}
\markboth{DRAFT}{Li \MakeLowercase{\textit{et al.}}: A Full Wave-Domain Linear MIMO-OFDM Receiver}

\maketitle

\begin{abstract}
This paper develops a wave-domain linear receiver for wideband MIMO-OFDM, where the main high-dimensional linear processing conventionally performed after sampling is physically shifted to propagation through stacked intelligent metasurfaces (SIMs). The proposed receiver transforms receive-port waveforms into OFDM-domain symbols through three cascaded SIM-enabled functional sections. The first section, SIM$_1$, performs channel-memory processing and channel shortening in an energy-concentration sense, suppressing the out-of-cyclic-prefix (CP) effective tail before OFDM analysis. SIM$_2$ realizes CP removal and discrete Fourier transform (DFT) analysis as cyclic-windowed OFDM matched filtering by exploiting useful-window memory and analytical Fourier phase coding. SIM$_3$ implements residual virtual-subcarrier MIMO linear detection, completing the wave-domain linear receive chain up to symbol slicing. Synchronization, control, slicing, and decoding remain digital, while the dominant linear receive operations are moved to the wave domain. A staged configuration framework is developed: the numerology-dependent OFDM analyzer is compiled offline, whereas the channel-dependent memory processor and detector are configured over the coherence time using architecture-aware learning-assisted control. Numerical results validate both functional replacement and link performance, confirming nearly lossless replacement under CP-sufficient MIMO-OFDM conditions, with robustness to residual timing offsets and effective channel-delay-profile perturbations. Substantial gains under CP-insufficient channels are also demonstrated by suppressing out-of-CP leakage before OFDM analysis.
\end{abstract}

\vspace{-0.5 cm}
\section{Introduction}
\label{sec:introduction}

Multiple-input multiple-output orthogonal frequency-division multiplexing (MIMO-OFDM) is a fundamental air-interface technology for wideband wireless communication \cite{OFDM,OFDM2}. A conventional MIMO-OFDM receiver first samples the incident receive signals through multiple radio-frequency (RF) chains and analog-to-digital converters (ADCs), and then performs cyclic-prefix (CP) removal, discrete Fourier transform (DFT)-based OFDM demodulation, and per-subcarrier MIMO detection in the digital baseband. This digital-baseband-first architecture is flexible and mature, but it also concentrates the receiver-side processing burden after high-dimensional sampling. With increasing array size, bandwidth, and carrier frequency in future 6G systems using millimeter-wave and terahertz bands \cite{vision1, vision2}, the cost of wideband RF chains, high-speed ADCs, DFT engines, and repeated subcarrier-domain matrix operations becomes increasingly significant. Hybrid analog-digital architectures and dynamic subarray structures reduce part of this burden by shifting spatial beamforming to the RF domain \cite{HY1,HY2,HY3}, but the essential OFDM receive chain is still largely executed in the digital baseband.

Stacked intelligent metasurfaces (SIMs) provide a different hardware paradigm for communication signal processing. By cascading multiple programmable metasurface layers, SIMs can implement controllable electromagnetic (EM) transformations directly in the wave domain with additional degrees of freedom (DoF) \cite{SIM0}. Early diffractive neural networks and diffractive metasurfaces demonstrated that multilayer EM structures can realize trainable analog transformations \cite{Lin2018D2NN,Liu2022D2NN}. Building on this principle, SIM-enabled wireless systems have been investigated for holographic MIMO transmission, wave-domain beamforming, radiation-pattern control, and channel shaping \cite{An2023SIM,Papazafeiropoulos_TWC_2024_HolographicSIM,Li1,Hassan_OJCOMS_2024_Beamforming,An_TWC_2025_MUBeamforming}. Recent studies have further extended SIMs to near-field communications, secure transmission, integrated sensing and communications, semantic communications, and cell-free networks \cite{Nadeem_WCNC_2024_ChannelEst,Yao_WCL_2024_ChannelEst,Papazafeiropoulos_WCL_2024_NearField,Niu_TIFS_2024_SecureSIM,Niu_WCL_2024_ISACSIM,Huang_WCL_2025_SemanticSIM,Li_TCOM_2024_CellFreeSIM}. Survey and in-wave computation studies have also emphasized that SIMs are not merely passive reflectors, but programmable multilayer EM processors capable of mapping signal-processing tasks into physical wave propagation \cite{survey1,survey2,NN,survey3}. These results establish SIMs as a promising platform for wave-domain computation and signal processing.

Despite these advances, most existing SIM-based communication architectures still use SIMs as auxiliary spatial processors. In receive-side systems, a SIM is typically configured as a memoryless analog combiner that mixes the received signal at the current time instant. Such a structure can reduce spatial dimension or suppress spatial interference, but it cannot access multiple delayed waveform samples and therefore cannot directly perform pre-DFT channel-memory processing. In wideband SIM-assisted MIMO-OFDM systems, multilayer metasurfaces have been shown to improve analog beamforming, subcarrier-domain spatial multiplexing, and multiuser resource interaction \cite{Li2,Li3,Li4, QLi}. However, the main OFDM receive operations, including channel-memory handling, CP-window selection, DFT analysis, and residual MIMO detection, are still performed after sampling in the digital baseband. Therefore, existing SIM-assisted receivers do not yet provide a physical realization of the receiver-side MIMO-OFDM linear processing chain.

This gap motivates a receiver-side question that is more fundamental than analog combining: can the main linear modules of a conventional MIMO-OFDM receiver be physically realized in the wave domain before high-dimensional digital demodulation? This question is the receiver-side counterpart of wave-domain OFDM synthesis at the transmitter. At the transmitter, the information symbols are known, and the device actively synthesizes the desired CP-OFDM waveform. The baseband-free wideband multiuser MIMO-OFDM transmitter was constructed via SIMs in \cite{Li5, Li6} to realize fully wave-domain symbol loading, precoding, and OFDM modulation, which can obtain comparable performance with fully digital architecture. However, at the receiver, the incoming waveform is unknown and distorted by frequency-selective multipath and noise \cite{DMA} and subject to timing uncertainty. Unlike transmitter-side synthesis, receiver-side analysis must identify the useful OFDM window from a distorted received waveform, so synchronization and channel memory cannot be separated from the wave-domain analyzer. Therefore, a wave-domain receiver must provide temporal memory for long-multipath processing, a synchronization-aware OFDM analyzer for CP-windowed frequency analysis, and a channel-dependent detector for residual stream separation.

To address this problem, this paper proposes a cascaded wave-domain linear MIMO-OFDM receiver enabled by delay-augmented SIM sections. For design clarity, the integrated cascaded SIM is decomposed into three functional sections, denoted by SIM$_1$, SIM$_2$, and SIM$_3$. This decomposition is introduced for modeling and optimization: the three sections may be physically regarded as cascaded regions of one multi-layer SIM architecture. The key distinction from end-to-end black-box wave-domain narrowband processors \cite{NBF} is that the proposed receiver does not learn a monolithic dense transformation from received samples to detected symbols. Instead, it follows the explainable structure of the MIMO-OFDM receive chain and assigns each functional section a receiver-specific role. SIM$_1$ upgrades a memoryless SIM into a spatio-temporal channel-memory processor by combining beamspace mapping, delay-augmented memory, and output recombination. SIM$_2$ materializes CP removal and DFT analysis as cyclic-windowed OFDM matched filtering by explicitly exploiting the analytical Fourier phase law of OFDM. SIM$_3$ realizes virtual-subcarrier residual linear detection over the OFDM-analysis interface. Synchronization/control, symbol slicing, and channel decoding remain digital, but the main high-dimensional linear receive operations are moved into the wave domain. This structured realization is important for both interpretability and hardware scalability. Moreover, to make the architecture implementable under finite hardware resources, we develop a staged configuration framework. SIM$_2$ depends only on the OFDM numerology and is compiled offline as a calibrated OFDM analyzer. SIM$_1$ and SIM$_3$ depend on the channel and are configured over the channel coherence time. Their optimization remains model-driven at the operator level, while architecture-aware learning-assisted controllers are used only to accelerate low-dimensional online configuration variables. Unlike learned controllers that directly generate black-box phase-shift matrices for SIM \cite{DRL,DRL2}, the proposed optimization method selects structured controls that generate a channel shortener and a residual detector consistent with the proposed wave-domain receiver architecture.

The main contributions are summarized as follows.
\begin{itemize}
\item We propose a receiver-side wave-domain realization principle for MIMO-OFDM, where the main high-dimensional linear receive chain, including channel-memory processing, CP-windowed DFT analysis, and residual MIMO detection, is physically redistributed from digital baseband to the cascaded SIM architecture.
\item We formulate SIM$_1$ as a delay-augmented spatio-temporal channel-memory processor. The resulting wave-domain finite impulse response (FIR) shortener suppresses the out-of-CP effective tail and concentrates the effective channel energy inside the CP-compatible window before OFDM analysis.
\item We develop SIM$_2$ as a cyclic-windowed OFDM-analysis materializer. Instead of separately implementing analog CP gating, serial-to-parallel conversion, and DFT, SIM$_2$ realizes the CP-windowed DFT projection through useful-window memory, Fourier phase-ramp coding, and wave-domain summation.
\item We construct SIM$_3$ as a unified virtual-subcarrier linear detector. Its block-level transfer structure is designed to materialize a colored-interference-aware MMSE-like detector in the wave domain while preserving the virtual OFDM-mode grouping recovered by SIM$_2$.
\item We provide a hardware-aware configuration and validation framework, including offline SIM$_2$ compilation, online SIM$_1$ and SIM$_3$ configuration, CP-insufficient and CP-sufficient performance evaluation, ablation studies, and robustness analysis with residual timing offsets and effective channel-delay-profile perturbations.
\end{itemize}

The rest of this paper is organized as follows. Section~\ref{sec:system_model} introduces the conventional receiver functions, signal model, and proposed SIM-enabled wave-domain architecture. Section~\ref{sec:receiver_modules} develops the three receiver modules and their design targets. Section~\ref{sec:solution_framework} presents the offline compilation and learning-assisted online configuration framework. Section~\ref{sec:results} provides numerical results and functional validation. Finally, Section~\ref{sec:conclusion} concludes the paper.

\vspace{-0.3 cm}
\section{Conventional MIMO-OFDM Receiver and Wave-Domain Architecture}
\label{sec:system_model}

This section establishes the architectural mapping and signal model used throughout the paper. We first decompose the conventional MIMO-OFDM receiver only into the linear modules that are relevant to this work, and then identify their SIM-enabled wave-domain counterparts.
\vspace{-0.5 cm}
\subsection{Functional Mapping and Signal Model}

\begin{figure*}
	\centerline{\includegraphics[width=0.8\textwidth]{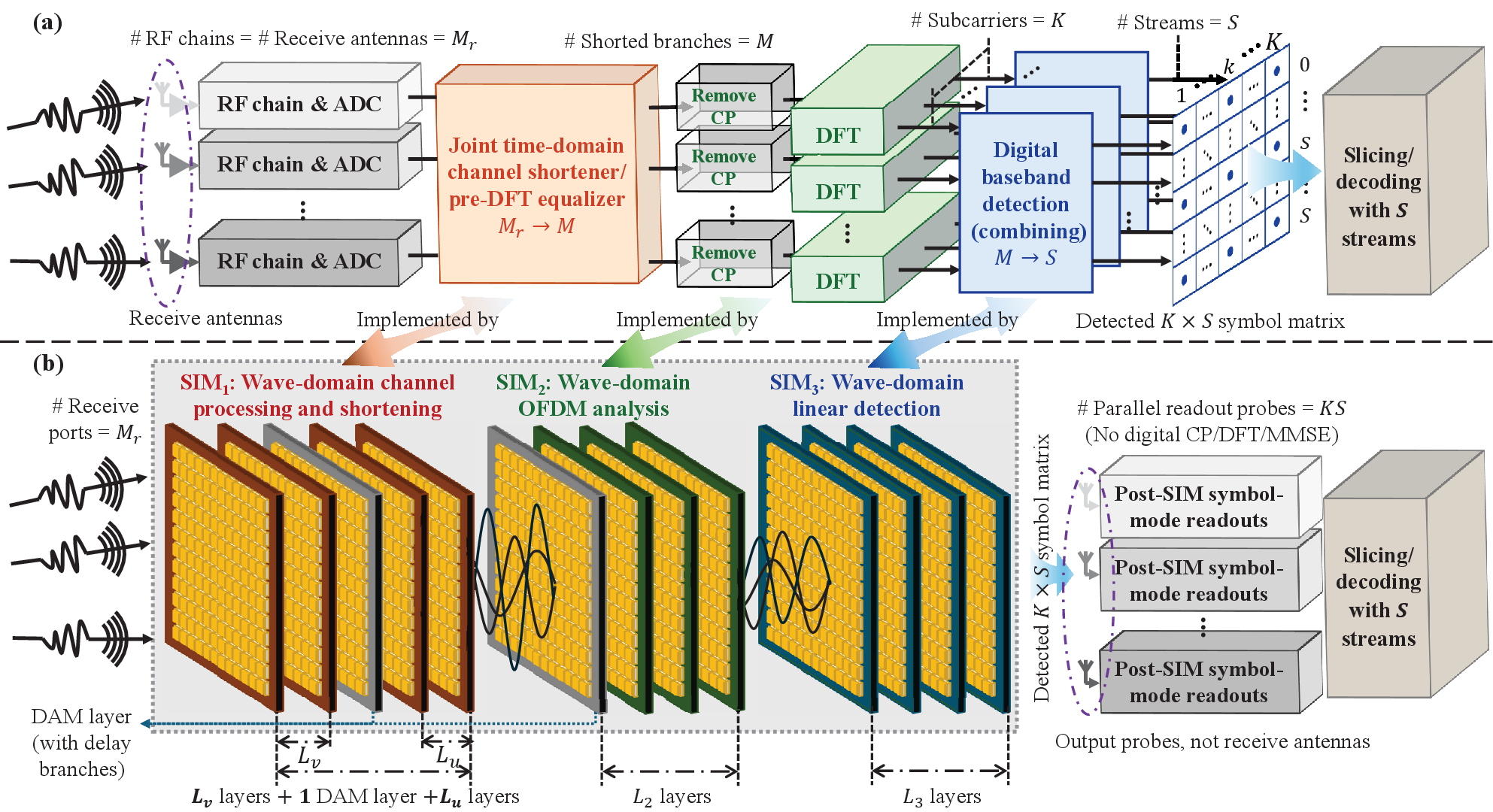}}
	\caption{Comparison of (a) the conventional fully digital wideband MIMO-OFDM receiver and (b) the proposed cascaded wave-domain linear receiver. The labels SIM$_1$, SIM$_2$, and SIM$_3$ denote functional sections for channel-memory processing, OFDM analysis, and residual linear detection, respectively; they may be implemented as cascaded regions of one integrated SIM architecture.}
	\label{fig:architecture}
\vspace{-0.5 cm}
\end{figure*}

As shown in Fig.~\ref{fig:architecture}, the conventional receiver performs CP removal, DFT, and subcarrier-wise MIMO detection after RF-chain/ADC sampling. The proposed receiver instead compiles these high-dimensional linear operations into the wave-domain realization using cascaded metasurfaces. For modeling and configuration clarity, the proposed receiver maps these same linear functions to three sections: SIM$_1$ performs delay-augmented pre-DFT channel-memory processing, SIM$_2$ materializes CP-windowed DFT analysis, and SIM$_3$ implements residual virtual-subcarrier detection. Their update timescales differ: SIM$_2$ is numerology-dependent and compiled offline, whereas SIM$_1$ and SIM$_3$ are channel-adaptive.

We consider a wideband MIMO-OFDM link with \(K\) subcarriers,
CP length \(N_{\rm CP}\), \(S\) transmitted streams, \(N_t\)
transmit ports, and \(M_r\) receive ports. The integer \(b\) denotes the OFDM block index, \(k\in\mathcal K=\{0,\ldots,K-1\}\) denotes the subcarrier index, \(m\in\{0,\ldots,K-1\}\) denotes the useful-sample index after CP removal, \(t\in\mathcal T=\{0,\ldots,T-1\}\) denotes the CP-extended sample index within one OFDM block, where \(T=K+N_{\rm CP}\), and \(n\in\mathbb Z\) denotes the global discrete-time sample index at sampling interval \(T_s\). The channel tap index is denoted by \(\ell\), while the delay-branch index of SIM$_1$ is denoted by \(p\). Let \(\mathbf{s}_b[k]\in\mathbb{C}^{S}\) denote the transmitted stream vector on subcarrier \(k\) of block \(b\), where
\(\mathbb{E}\{\mathbf{s}_b[k]\mathbf{s}_b^{H}[k]\}=\mathbf{I}_{S}\). The stacked frequency-domain vector over one OFDM block is
\begin{align}
\mathbf{s}_b
=
\left[
\mathbf{s}_b^{T}[0],
\mathbf{s}_b^{T}[1],
\ldots,
\mathbf{s}_b^{T}[K-1]
\right]^{T}
\in\mathbb{C}^{KS}.
\end{align}

The transmitter-side OFDM synthesis, including subcarrier-domain
precoding, IDFT, and CP insertion, is represented by a block operator
\(\mathbf{Q}_{\rm TX}\in\mathbb C^{TN_t\times KS}\). Therefore, the transmitted time-domain block can be written as
\begin{align}
\mathbf{x}_b&=\mathbf{Q}_{\rm TX}\mathbf{s}_b \in\mathbb C^{TN_t} \nonumber \\
&=
\left[
\mathbf x^T[n_b],
\mathbf x^T[n_b+1],
\ldots,
\mathbf x^T[n_b+T-1]
\right]^T,
\end{align}
where \(\mathbf x[n]\in\mathbb C^{N_t}\) is the transmit-port sample vector at global time index \(n\). The nominal start of block \(b\) is \(n_b=n_0+bT\), where \(n_0\) is the coarse timing reference. For notational simplicity, \(n_0=0\) can be assumed after synchronization.

At the receiver, the instantaneous receive-port vector is
\begin{align}
\mathbf r[n]
=
\left[
r_1[n],r_2[n],\ldots,r_{M_r}[n]
\right]^T
\in\mathbb C^{M_r},
\end{align}
where \(r_i[n]\) is the synchronized complex-baseband equivalent of the EM field coupled to the \(i\)-th receive port at time \(nT_s\). 

Let \(\mathbf{H}[\ell]\in\mathbb{C}^{M_r\times N_t}\), \(\ell=0,\ldots,L_h-1\), denote the discrete-time wideband MIMO channel taps between the \(N_t\) transmit ports and the \(M_r\) receive ports. The integer \(L_h\) is the channel length in samples, so the physical delay spread is approximately \((L_h-1)T_s\). The main derivation uses the synchronized equivalent baseband model after coarse timing acquisition and residual carrier-frequency offset (CFO) compensation by a low-dimensional control branch:
\begin{align}
\mathbf{r}[n]
=
\sum_{\ell=0}^{L_h-1}
\mathbf{H}[\ell]\mathbf{x}[n-\ell]
+
\mathbf{w}[n] \in\mathbb C^{M_r},
\label{eq:synchronized_rx_model}
\end{align}
where \(\mathbf{x}[n-\ell]\in\mathbb C^{N_t}\) is the transmit-port sample vector delayed by \(\ell\) samples, and \(\mathbf{w}[n]\sim\mathcal{CN}(\mathbf{0},\sigma^2\mathbf{I}_{M_r})\) denotes receive-port noise.

For the \(b\)-th CP-extended OFDM receive block, the vector \(\mathbf r_b\in \mathbb{C}^{TM_r}\) stacks the \(T\) receive-port vectors associated with the nominal block interval \([n_b,n_b+T-1]\):
\begin{align}
\mathbf{r}_b
=
\left[
\mathbf{r}^{T}[n_b],
\mathbf{r}^{T}[n_b+1],
\ldots,
\mathbf{r}^{T}[n_b+T-1]
\right]^T.
\end{align}

We define the augmented block \(\bar{\mathbf{r}}_b \in \mathbb{C}^{(T+P_1-1)M_r}\)
\begin{align}
\bar{\mathbf{r}}_b
=
\left[
\mathbf{r}^{T}[n_b-P_1+1],
\ldots,
\mathbf{r}^{T}[n_b+T-1]
\right]^T,
\end{align}
which consists of the CP-extended block \(\mathbf r_b\) and the historical receive-port samples required by the causal wave-domain FIR.

Let \(\bar{\mathbf{H}}_{b,b}\in\mathbb C^{(T+P_1-1)M_r\times TN_t}\) denote the block Toeplitz convolution matrix from the current transmitted CP-extended block to \(\bar{\mathbf r}_b\), and let \(\bar{\mathbf{H}}_{b,b-1}\in\mathbb C^{(T+P_1-1)M_r\times TN_t}\) denote the corresponding leakage matrix from the previous block. Then
\begin{align}
\bar{\mathbf{r}}_b
=
\bar{\mathbf{H}}_{b,b}\mathbf{Q}_{\rm TX}\mathbf{s}_b
+
\bar{\mathbf{H}}_{b,b-1}\mathbf{Q}_{\rm TX}\mathbf{s}_{b-1}
+
\bar{\mathbf{w}}_b,
\label{eq:block_conv_with_ibi}
\end{align}
where \(\bar{\mathbf w}_b\in\mathbb C^{(T+P_1-1)M_r}\) is the stacked noise vector. The second term represents possible inter-block interference (IBI) caused by the channel memory and by the historical samples required by the causal SIM$_1$ input. When an isolated-block evaluation or a sufficiently long guard interval is used, this term vanishes. In the continuous-block operation, it is retained as residual IBI and can be included in the interference covariance. For compact notation, the following derivation focuses on the desired current-block contribution and writes
\(\bar{\mathbf H}_{\rm conv}\triangleq\bar{\mathbf H}_{b,b}\), while treating the previous-block term as part of the residual interference when present.

\begin{remark}
\textnormal{The relation between \(N_{\rm CP}\) and \(L_h\) determines the role of SIM$_1$. If $L_h-1\leq N_{\rm CP}$, then the standard CP-OFDM circular-convolution condition is satisfied after CP removal. In this CP-sufficient regime, a conventional digital DFT+MMSE receiver is already well matched to the channel, and the proposed wave-domain linear receiver aims to reproduce it with negligible functional-replacement loss. If $L_h-1>N_{\rm CP}$, then conventional digital DFT demodulation suffers from out-of-CP leakage and residual inter-carrier interference (ICI). This CP-insufficient regime exposes the intended role of SIM$_1$: the effective channel energy should be concentrated inside the CP-compatible window before OFDM analysis.}
\end{remark}
\vspace{-0.5cm}

\subsection{Hardware Model and Control Assumptions}

We first introduce the standard phase-only SIM core, and then summarize the delay-memory extension required by SIM$_1$ and SIM$_2$. A standard SIM core consists of multiple metasurface layers separated by free-space, dielectric, waveguide, or other known coupling regions \cite{Liu2022D2NN}. For the \(i\)-th SIM core, let \(N_i^{\rm in}\) and \(N_i^{\rm out}\) denote the numbers of input-interface and output-interface elements, respectively. These interface elements can be implemented as boundary meta-atoms, aperture-coupled elements, or readout probes. Let \(L_i\) denote the number of internal metasurface layers used for wave-domain mixing, and let \(d_{i,l}\) denote the number of meta-atoms at internal layer \(l\in\mathcal{L}_i=\{1,\ldots,L_i\}\). The fixed coupling matrix \(\mathbf W_i^1\in\mathbb C^{d_{i,1}\times N_i^{\rm in}}\) maps the input interface to the first internal layer, while \(\mathbf W_i^l\in\mathbb C^{d_{i,l}\times d_{i,l-1}}\), \(l\ge2\), describes propagation from layer \(l-1\) to layer \(l\). The readout matrix \(\mathbf S_i\in\mathbb C^{N_i^{\rm out}\times d_{i,L_i}}\) maps the last internal layer to the output interface. The phase response of layer \(l\) is
\begin{equation}
\boldsymbol{\Psi}_i^l
=
{\rm diag}
\left(
e^{j\theta_{i,1}^l},
e^{j\theta_{i,2}^l},
\ldots,
e^{j\theta_{i,d_{i,l}}^l}
\right).
\end{equation}

For \(B_{\theta,i}\)-bit quantized phase control, \(\theta_{i,m}^l\in\mathcal B_i\), where
\(\mathcal B_i=\{0,2\pi/2^{B_{\theta,i}},\ldots,2\pi(2^{B_{\theta,i}}-1)/2^{B_{\theta,i}}\}\). The propagation matrix \(\mathbf W_i^l\) is determined by the EM coupling between the previous interface/layer and the current layer. For the \(i\)-th SIM, its \((m,m')\)-th entry is described by a Rayleigh--Sommerfeld-type response
\begin{equation}
[\mathbf W_i^{l}]_{m,m'}
=
\frac{A_i\delta_i}{r_{i,l,m,m'}^2}
\left(
\frac{1}{2\pi r_{i,l,m,m'}}
-
j\frac{f_c}{c}
\right)
e^{j2\pi f_c r_{i,l,m,m'}/c},
    \label{eq:rs_propagation}
\end{equation}
where \(A_i\) is the meta-atom area, \(\delta_i\) is the inter-layer spacing, \(r_{i,l,m,m'}\) is the propagation distance between the corresponding elements, \(f_c\) is the carrier frequency, and \(c\) is the speed of light. For \(l=1\), \(m'\) indexes the input-interface elements; for \(l\ge2\), it indexes the meta-atoms on layer \(l-1\).

\begin{table}[t]
\centering
\caption{Delay-Memory Capabilities Used by the Proposed Receiver}
\label{tab:delay_hardware_evidence}
\footnotesize
\setlength{\tabcolsep}{2.5pt}
\renewcommand{\arraystretch}{1.18}
\begin{tabular}{@{}>{\raggedright\arraybackslash}p{1.75cm}
>{\raggedright\arraybackslash}p{2.5cm}
>{\raggedright\arraybackslash}p{2.9cm}
>{\centering\arraybackslash}p{1.25cm}@{}}
\toprule
Capability & Physical realization & Role in this work & Refs. \\
\midrule
Coherent storage \& retrieval
& EIT meta-atoms switch between radiative and trapped states.
& Provides phase-coherent memory for delayed waveform samples.
& \cite{Nakanishi2013EITStorage,Nakanishi2018EITRetrieval} \\
\addlinespace[1pt]
Programmable holding time
& Control timing determines the field release instant.
& Forms SIM$_1$ tapped branches and SIM$_2$ useful-window branches.
& \cite{Nakanishi2018EITRetrieval,An2024AdjustableDelayRIS} \\
\bottomrule
\end{tabular}
\setlength{\tabcolsep}{6pt}
\vspace{-0.6cm}
\end{table}

\begin{table*}
\centering
\caption{Wave-Domain Operators and Hardware Dimensions of the Cascaded Receiver Sections}
\label{tab:sim_hardware_dimensions}
\footnotesize
\renewcommand{\arraystretch}{1.12}
\begin{tabular}{p{2.60cm}p{2.85cm}p{3.5cm}p{2.4cm}p{4.6cm}}
\hline
Section & Wave-domain map & Boundary interface & Internal layers & Configurable / optimized DoF \\
\hline
SIM$_1$ input core \(\mathbf V\)
& \(\mathbb C^{M_r}\!\rightarrow\!\mathbb C^{R_{\rm c}}\)
& \(M_r\) receive elements; \(R_{\rm c}\) beamspace readouts
& \(L_V\) layers with \(d_{V,l}\) meta-atoms per layer
& \(\boldsymbol\Omega_V=\{\boldsymbol\Psi_V^l\}_{l=1}^{L_V}\) \\
SIM$_1$ delay-tap core
& \(\mathbf C(z):\mathbb C^{R_{\rm c}}\!\rightarrow\!\mathbb C^{R_{\rm c}}\)
& \(R_{\rm c}\) beamspace branches
& One DAM layer
& \(\{\tau_p=pT_s,\boldsymbol\Lambda_p\}_{p=0}^{P_1-1}\), \(\boldsymbol\Lambda_p\in\mathbb C^{R_{\rm c}\times R_{\rm c}}\) \\
SIM$_1$ output core \(\mathbf U\)
& \(\mathbb C^{R_{\rm c}}\!\rightarrow\!\mathbb C^{M}\)
& \(R_{\rm c}\) beamspace inputs; \(M\) shortened outputs
& \(L_U\) layers with \(d_{U,l}\) meta-atoms per layer
& \(\boldsymbol\Omega_U=\{\boldsymbol\Psi_U^l\}_{l=1}^{L_U}\) \\
SIM$_2$ OFDM analyzer
& \(\mathcal Q_{\rm OA}:\mathbb C^{TM}\!\rightarrow\!\mathbb C^{KM}\)
& Useful-window \(TM\!\rightarrow\!KM\); \(K\) groups of size \(M\)
& \(L_2\) layers with \(d_{2,l}\) meta-atoms per layer
& \(\boldsymbol\Omega_2=\{\boldsymbol\Psi_2^l\}_{l=1}^{L_2}\); useful-window delays; \(\{\mathbf D_k\}_{k=0}^{K-1}\) Fourier phases \\
SIM$_3$ detector section
& \(\mathcal Q_{\rm DET}:\mathbb C^{KM}\!\rightarrow\!\mathbb C^{KS}\)
& \(K\) diagonal blocks of size \(S\times M\)
& \(L_3\) layers with \(d_{3,l}\) meta-atoms per layer
& \(\boldsymbol\Omega_3=\{\boldsymbol\Psi_3^l\}_{l=1}^{L_3}\); virtual-subcarrier block structure; \(\{\alpha_{3,k}\}_{k=0}^{K-1}\) calibration \\
\hline
\end{tabular}
\vspace{-0.6cm}
\end{table*}

The standard SIM core above is memoryless with respect to the discrete-time waveform. Since SIM$_1$ requires tapped-delay channel-memory processing and SIM$_2$ requires useful-window OFDM sample exposure, the proposed receiver uses calibrated delay-augmented metasurface (DAM) layers/interfaces to provide temporal memory. Prior varactor-loaded electromagnetically induced transparency (EIT) meta-atoms have demonstrated phase-coherent EM-wave storage and retrieval, while adjustable-delay metasurfaces translate this mechanism into controllable delay DoF for OFDM systems \cite{Nakanishi2013EITStorage,Nakanishi2018EITRetrieval,An2024AdjustableDelayRIS}. In this paper, the DAM layer is not treated as a separate receiver architecture, but it is a temporal-memory extension embedded into the SIM sections. Table~\ref{tab:delay_hardware_evidence} summarizes only the DAM capabilities used by the proposed design. Circuit-level storage loss, switching error, and delay-line calibration are hardware-specific issues and are not explicitly modeled in this work; their aggregate effects can be absorbed into fitted calibration factors and evaluated in prototype-level studies.

The main hardware dimensions are summarized in Table~\ref{tab:sim_hardware_dimensions}. The boundary-interface sizes specify the number of signal-carrying receive, branch, or readout elements, whereas \(d_{i,l}\) specifies the number of programmable meta-atoms in internal mixing layer \(l\). The last column lists the phase, delay, tap, or calibration DoF configured by the proposed design.
\vspace{-0.2cm}
\section{Wave-Domain Receiver Modules and Design}
\label{sec:receiver_modules}

This section develops the three functional sections of the cascaded receiver. Each subsection states what the corresponding SIM section physically performs, why the operation is required, and what configuration target it should realize. Detailed numerical updates and learning-assisted configuration steps are deferred to Section~\ref{sec:solution_framework}.
\vspace{-0.5cm}
\subsection{SIM$_1$: Delay-Augmented Channel Shortening}
\label{subsec:sim1_module}

In contrast to a memoryless SIM that can only process the instantaneous receive-port vector \(\mathbf r[n]\), SIM$_1$ jointly processes the current and historical receive-port samples, thereby realizing a matrix-valued FIR operation in the wave domain. In this paper, channel shortening refers to effective delay-domain energy concentration, not physical support reduction. Since an FIR shortener can extend the algebraic support to \(L_h+P_1-1\), the design objective is to suppress the out-of-CP effective tail and preserve the useful in-CP energy. 

\subsubsection{Architecture and Operator Model}
SIM$_1$ is organized as an input beamspace core, a weighted tapped-delay DAM layer, and an output recombination core. The receive-port vector is first mapped to an \(R_{\rm c}\)-dimensional beamspace,
\begin{equation}
\mathbf b[n]=\mathbf V(\boldsymbol\Omega_V)\mathbf r[n]\in\mathbb C^{R_{\rm c}},
\end{equation}
where \(R_{\rm c}\) is the number of retained beamspace modes. We denote the SIM$_1$ delay-branch index set by \(\mathcal P_1=\{0,\ldots,P_1-1\}\). The input beamspace core is realized by
\begin{equation}
\mathbf V (\boldsymbol\Omega_V)
=
\mathbf S_V
\boldsymbol{\Psi}_{V}^{L_V}
\mathbf W_V^{L_V}
\cdots
\boldsymbol{\Psi}_{V}^{1}
\mathbf W_V^{1}
\in\mathbb C^{R_{\rm c}\times M_r},
\end{equation}
where the phase configuration is \(\boldsymbol\Omega_V=\{\boldsymbol{\Psi}_V^l\}_{l=1}^{L_V}\), \(\boldsymbol{\Psi}_{V}^{l}\) is the phase-shift matrix of the \(l\)-th layer, \(\mathbf W_V^{l}\) follows the common SIM coupling convention, and \(\mathbf S_V\in\mathbb C^{R_{\rm c}\times d_{V,L_V}}\) reads out the beamspace interface.

\begin{figure}[t]
\centering
\includegraphics[width=0.4\textwidth]{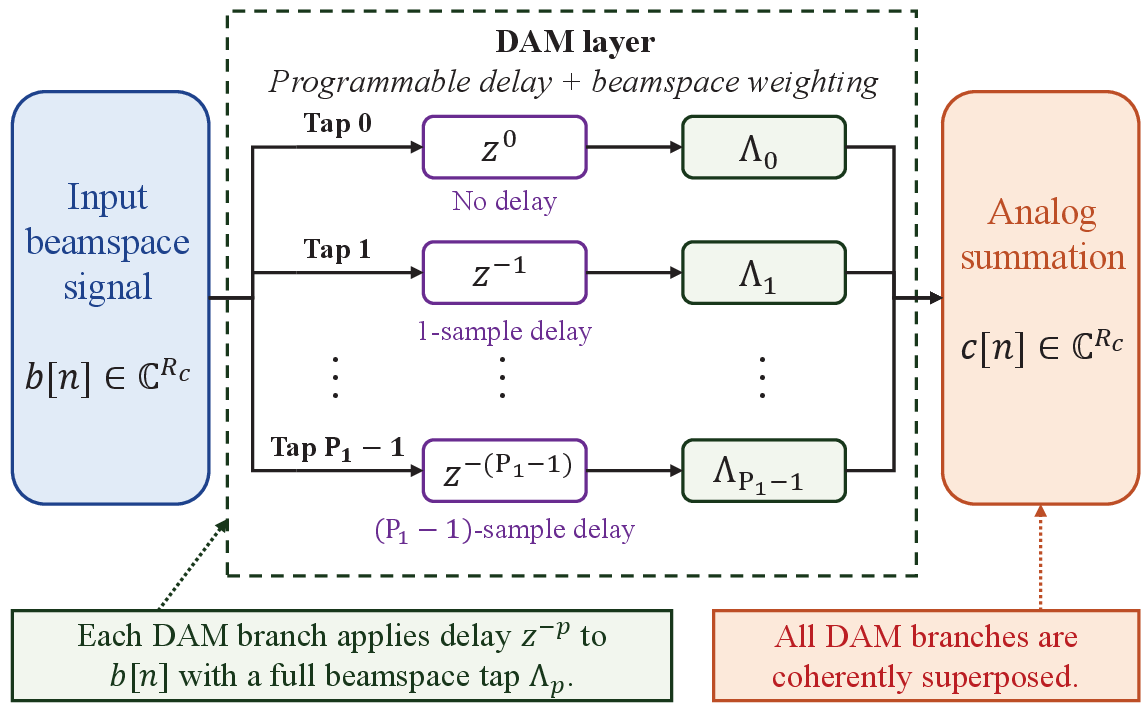}
\caption{Signal-flow interpretation of the weighted tapped-delay DAM layer in SIM$_1$.}
\label{fig:sim1_dam_signal_flow}
\vspace{-0.5cm}
\end{figure}

The DAM layer then applies \(P_1\) sample-spaced delay branches and tap-dependent weights $\boldsymbol{\Lambda}_p\in\mathbb C^{R_{\rm c}\times R_{\rm c}}$:
\begin{align}
\mathbf c[n]
=
\sum_{p=0}^{P_1-1}
\boldsymbol{\Lambda}_p\mathbf b[n-p] \in\mathbb C^{R_{\rm c}},
\label{eq:sim1_dam_compact}
\end{align}
The signal-flow interpretation is shown in Fig.~\ref{fig:sim1_dam_signal_flow}: the \(p\)-th branch applies a delay of \(pT_s\), the matrix \(\boldsymbol{\Lambda}_p\), and coherent analog recombination. In the performance-oriented implementation considered here, \(\boldsymbol\Lambda_p=[\lambda_{p,i,j}]_{i,j=1}^{R_{\rm c}}\) is a full low-dimensional beamspace tap with \(\lambda_{p,i,j}=\rho_{p,i,j}e^{j\varphi_{p,i,j}}\), \(0\le\rho_{p,i,j}\le\rho_{\max}\), and \(\varphi_{p,i,j}\in[0,2\pi)\), where \(\rho_{\max}\) is the maximum calibrated tap magnitude allowed by the delay-branch loading circuit.

The delayed beamspace waveform is finally recombined into an \(M\)-element shortened-output interface:
\begin{align}
\mathbf u[n]
&=
\mathbf U(\boldsymbol\Omega_U)\mathbf c[n] \in\mathbb C^{M}, \\
\mathbf U(\boldsymbol\Omega_U)
&=
\mathbf S_U
\boldsymbol{\Psi}_{U}^{L_U}
\mathbf W_U^{L_U}
\cdots
\boldsymbol{\Psi}_{U}^{1}
\mathbf W_U^{1}
\in\mathbb C^{M\times R_{\rm c}},
\end{align}
where \(\mathbf S_U\in\mathbb C^{M\times d_{U,L_U}}\). The definitions of \(\boldsymbol\Omega_U\), \(\boldsymbol{\Psi}_{U}^{l}\), and \(\mathbf W_U^{l}\) follow the same layered-SIM convention as the input core.

Combining the three blocks gives the structured SIM$_1$ tap
\begin{align}
\mathbf Q_p^{\rm CS}(\boldsymbol\Omega_1)
=
\mathbf U(\boldsymbol\Omega_{U})
\boldsymbol\Lambda_p
\mathbf V(\boldsymbol\Omega_{V})
\in\mathbb C^{M\times M_r},
\label{eq:sim1_structured_tap_constraint}
\end{align}
where \(\boldsymbol\Omega_1=\{\boldsymbol\Omega_U,\boldsymbol\Omega_V,\boldsymbol\Lambda_0,\ldots,\boldsymbol\Lambda_{P_1-1}\}\).  The layer sizes of \(\mathbf V\) and \(\mathbf U\) are specified in Table~\ref{tab:sim_hardware_dimensions}. Thus, at each global index \(n\), SIM$_1$ forms the delay-lifted receive-port vector \(\bar{\mathbf r}[n]\in\mathbb C^{P_1M_r}\),
\begin{align}
\bar{\mathbf r}[n]
=
\left[
\mathbf r^T[n],
\mathbf r^T[n-1],
\ldots,
\mathbf r^T[n-P_1+1]
\right]^T.
\end{align}

With \(\mathbf Q_{\rm CS}(\boldsymbol\Omega_1)=[\mathbf Q_0^{\rm CS},\ldots,\mathbf Q_{P_1-1}^{\rm CS}]\), the output is
\begin{align}
\mathbf u[n]
=
\mathbf Q_{\rm CS}(\boldsymbol\Omega_1)
\bar{\mathbf r}[n]
=
\sum_{p=0}^{P_1-1}
\mathbf Q_p^{\rm CS}\mathbf r[n-p].
\label{eq:sim1_sample_fir}
\end{align}

For block-level OFDM analysis, stacking \(T\) equations yields \(\mathbf u_b=\mathcal Q_{\rm CS}\bar{\mathbf r}_b\), where \(\bar{\mathbf r}_b\) collects the receive-port samples needed by the causal FIR shortener. The induced block operator \(\mathcal Q_{\rm CS}\in\mathbb C^{TM\times (T+P_1-1)M_r}\) has the Toeplitz form
\begin{align}
\mathcal Q_{\rm CS}
=
\begin{bmatrix}
\mathbf Q_{P_1-1}^{\rm CS}
& \cdots
& \mathbf Q_0^{\rm CS}
& \mathbf 0
& \cdots
& \mathbf 0
\\
\mathbf 0
& \mathbf Q_{P_1-1}^{\rm CS}
& \cdots
& \mathbf Q_0^{\rm CS}
& \ddots
& \vdots
\\
\vdots
& \ddots
& \ddots
& \ddots
& \ddots
& \mathbf 0
\\
\mathbf 0
& \cdots
& \mathbf 0
& \mathbf Q_{P_1-1}^{\rm CS}
& \cdots
& \mathbf Q_0^{\rm CS}
\end{bmatrix},
\label{eq:sim1_block_toeplitz}
\end{align}
where each nonzero block in \(\mathcal Q_{\rm CS}\) has dimension \(M\times M_r\).

\subsubsection{Configuration Target}

Given the structured FIR taps in \eqref{eq:sim1_sample_fir}, the effective channel after SIM$_1$ is
\begin{align}
\mathbf G[d]
=
\sum_{p=0}^{P_1-1}
\mathbf Q_p^{\rm CS}\mathbf H[d-p],
\label{eq:sim1_effectivechannel}
\end{align}
where \(\mathbf{H}[d-p]=\mathbf{0}\) outside the valid channel support. Let \(L_{\rm eff}=L_h+P_1-1\) denote the length of the effective FIR channel. To avoid the trivial solution that switches off the receiver, SIM$_1$ is designed to suppress the out-of-CP residual energy while preserving useful effective-channel energy inside the CP-compatible window. Define
\begin{align}
\mathcal E_{\rm in}
&=
\sum_{d=0}^{\min\{N_{\rm CP},L_{\rm eff}-1\}}
\|\mathbf G[d]\|_F^2,
\nonumber\\
\mathcal E_{\rm out}
&=
\sum_{d=N_{\rm CP}+1}^{L_{\rm eff}-1}
\|\mathbf G[d]\|_F^2,
\label{eq:cs_energy_terms}
\end{align}
where \(\mathcal E_{\rm out}=0\) if \(N_{\rm CP}\ge L_{\rm eff}-1\). The normalized out-of-CP tail ratio and FIR noise-enhancement proxy are
\begin{align}
\mathcal L_{\rm tail}
&=
\frac{\mathcal E_{\rm out}}{\mathcal E_{\rm in}+\epsilon_0},
\label{eq:tail_loss_refined}\\
\mathcal L_{\rm noise}^{(1)}
&=
\frac{\sigma^2}{M}
\sum_{p=0}^{P_1-1}
\left\|
\mathbf Q_p^{\rm CS}
\right\|_F^2,
\label{eq:sim1_noise_proxy}
\end{align}
where \(\epsilon_0>0\) is a small constant for numerical stability.

The channel-adaptive SIM$_1$ design is therefore formulated as a shortened-SINR-oriented energy-concentration problem:
\begin{subequations}
\begin{align}
(\mathcal P_{\rm CS}):~
\min_{\boldsymbol\Omega_{U},\boldsymbol\Omega_{V},\{\boldsymbol\Lambda_p\}}\quad&
\mathcal L_{\rm CS}
=
\frac{
\mathcal E_{\rm out}
+
\lambda_{\rm reg}\mathcal L_{\rm noise}^{(1)}
}{
\mathcal E_{\rm in}+\epsilon_0
}
\\
{\rm s.t.}\quad&
\mathcal E_{\rm in}\ge \gamma_{\rm sig},
\label{eq:cs_signal_constraint}\\
&
\mathbf U(\boldsymbol\Omega_{U})\in\mathcal C_U,\\
&
\mathbf V(\boldsymbol\Omega_{V})\in\mathcal C_V,\\
&
\boldsymbol\Lambda_p\in\mathcal C_\Lambda,\quad p\in\mathcal{P}_1.
\end{align}
\label{eq:stage2_cs}
\end{subequations}
\!\!\!where \(\lambda_{\rm reg}\ge0\) controls the regularization against noise enhancement and \(\gamma_{\rm sig}>0\) is the minimum useful effective-channel energy that must be preserved inside the CP-compatible window. The constraint in \eqref{eq:cs_signal_constraint} removes the all-zero shortener from the feasible set. In the normalized implementation used in the simulations, this useful-energy requirement is enforced by the nonzero CP-compatible target in the reference-shortener generation, rather than tuned as an additional performance hyperparameter. \(\mathcal C_U\) and \(\mathcal C_V\) are the phase-only layered SIM feasible sets of the input and output SIM cores, while \(\mathcal C_\Lambda\) specifies the beamspace tap constraints, including the magnitude bound \(\rho_{\max}\). In the online implementation, \(\lambda_{\rm reg}\) is instantiated by \(\lambda_{\rm reg}^{\rm eff}\) in Section~\ref{sec:solution_framework}, together with a tail-emphasis level used in the weighted target-generation step. Rate and ICI are used for controller reward/evaluation rather than being additional independent terms in \((\mathcal P_{\rm CS})\).

\vspace{-0.5cm}
\subsection{SIM$_2$: CP-Windowed OFDM Analysis}
\label{subsec:sim2_module}

SIM$_2$ materializes CP removal and DFT analysis as a wave-domain OFDM analyzer. The key observation is that, after SIM$_1$ produces the shortened waveform \(\mathbf u[n]\), the \(k\)-th OFDM output is a cyclic-windowed matched-filter projection:
\begin{equation}
    \mathbf y_b[k]
    =
    \sum_{m=0}^{K-1}
    \frac{1}{\sqrt K}
    e^{-j2\pi km/K}
    \mathbf u[n_b+N_{\rm CP}+m].
    \label{eq:subcarrier_projection}
\end{equation}

\subsubsection{Architecture and Operator Model}
Rather than fitting a dense CP-removal-and-DFT matrix, SIM$_2$ separates the known OFDM structure into useful-window memory, Fourier phase coding, and a shared wave-domain summation core. In the synchronized case, the useful-window memory exposes the \(K\) post-CP samples at one readout instant:
\begin{align}
\mathbf u_{b,{\rm use}}
&=
\left[
\mathbf u^{T}[n_b+N_{\rm CP}],
\ldots,
\mathbf u^{T}[n_b+N_{\rm CP}+K-1]
\right]^T \nonumber\\
&=
(\mathbf R_{\rm CP}\otimes\mathbf I_{M})\mathbf u_b
\in\mathbb C^{KM},
\end{align}
where \(\mathbf R_{\rm CP}=[\mathbf 0_{K\times N_{\rm CP}}\ \mathbf I_K]\) and $\otimes$ is Kronecker product. The CP samples are excluded by the cyclic window before DFT-like accumulation. A DAM-based storage-release layer can implement this operation by delaying the \(m\)-th useful branch by approximately \((K-1-m)T_s\), so that all useful samples are simultaneously available after the last useful sample arrives.

For the \(k\)-th virtual subcarrier group, the analytical Fourier phase-ramp matrix is
\begin{align}
\mathbf D_k
=
\operatorname{diag}
\left(
1,
e^{-j2\pi k/K},
\ldots,
e^{-j2\pi k(K-1)/K}
\right).
\end{align}

The corresponding Fourier-coded useful-window vector is
\begin{align}
\tilde{\mathbf u}_{b,k}
=
(\mathbf D_k\otimes\mathbf I_{M})\mathbf u_{b,{\rm use}}
\in\mathbb C^{KM}.
\end{align}

Since these Fourier phase matrices are fixed by the OFDM numerology, SIM$_2$ does not require all internal layers to be time-varying. The only fitted wave-domain component is the shared summation core
\begin{align}
\mathbf C_{\rm sum}(\boldsymbol\Omega_2)
=
\mathbf S_2
\boldsymbol{\Psi}_2^{L_2}\mathbf W_2^{L_2}
\cdots
\boldsymbol{\Psi}_2^{1}\mathbf W_2^{1}
\in\mathbb C^{M\times KM},
\end{align}
where the phase configuration is \(\boldsymbol\Omega_2=\{\boldsymbol{\Psi}_2^l\}_{l=1}^{L_2}\), \(\boldsymbol{\Psi}_{2}^{l}\) is the phase-shift matrix of the \(l\)-th layer, \(\mathbf W_2^{l}\) follows the common SIM coupling convention, and \(\mathbf S_2\in\mathbb C^{M\times d_{2,L_2}}\) reads out one \(M\)-element virtual-subcarrier group. The \(k\)-th readout is
\begin{align}
\mathbf y_b[k]
=
\mathbf C_{\rm sum}(\boldsymbol\Omega_2)
(\mathbf D_k\otimes\mathbf I_{M})
\mathbf u_{b,{\rm use}}
\in\mathbb C^{M},
\end{align}
and the complete virtual OFDM-analysis output is \(\mathbf y_b=[\mathbf y_b^T[0],\ldots,\mathbf y_b^T[K-1]]^T\in\mathbb C^{KM}\). Stacking all virtual subcarrier groups gives
\begin{align}
\mathcal Q_{\rm OA}(\boldsymbol\Omega_2)
=
\begin{bmatrix}
\mathbf C_{\rm sum}(\boldsymbol\Omega_2)
(\mathbf D_0\otimes\mathbf I_{M})\\
\mathbf C_{\rm sum}(\boldsymbol\Omega_2)
(\mathbf D_1\otimes\mathbf I_{M})\\
\vdots\\
\mathbf C_{\rm sum}(\boldsymbol\Omega_2)
(\mathbf D_{K-1}\otimes\mathbf I_{M})
\end{bmatrix}
(\mathbf R_{\rm CP}\otimes\mathbf I_{M}).
\label{eq:sim2_stacked_oa}
\end{align}

In the fully parallel implementation, the useful-window interface feeds \(K\) Fourier-coded branches and \(K\) readout groups. A time-reused implementation can instead reuse the same summation core over \(k\), trading latency for hardware area. In both cases, \(\mathbf C_{\rm sum}\) is shared and only the analytical phase ramp changes across virtual subcarrier groups.

\subsubsection{Configuration Target}

SIM$_2$ depends only on the OFDM numerology \((K,N_{\rm CP})\) and \(M\), so it is compiled offline. The ideal CP-windowed DFT operator is
\begin{equation}
\mathcal Q_{\rm OA}^{\star}
=
(\bF_K\otimes\bI_{M})
(\bR_{\rm CP}\otimes\bI_{M}),
\label{eq:ideal_oa}
\end{equation}
where \(\bF_K\) is the normalized \(K\)-point DFT matrix. Since \(\bR_{\rm CP}\) and \(\{\mathbf D_k\}\) are analytically fixed, the fitting only targets the summation core
\begin{align}
\mathbf C_{\rm sum}^{\star}
=
\frac{1}{\sqrt K}
(\mathbf 1_K^T\otimes\mathbf I_{M}).
\end{align}

The offline compilation problem is
\begin{subequations}
\begin{align}
(\mathcal P_{\rm OA}):~\min_{\boldsymbol\Omega_2,\alpha_2}\quad&
\mathcal L_{\rm OA}=
\frac{
\left\|
\alpha_2
\mathbf C_{\rm sum}(\boldsymbol\Omega_2)
-
\mathbf C_{\rm sum}^{\star}
\right\|_F^2
}{
\left\|
\mathbf C_{\rm sum}^{\star}
\right\|_F^2
}. \\
{\rm s.t.}\quad&
\theta_{2,m}^l\in \mathcal B_2,\quad \forall l,m,
\end{align}
\label{prob:oa_fit_sec3}
\end{subequations}
\!\!\!where \(\alpha_2\in\mathbb C\) is a scalar calibration factor. This fitting loss is the primary offline compilation objective for SIM$_2$. The mode orthogonality of the complete analyzer is reported in the numerical section by the row-Gram off-diagonal-to-diagonal ratio as a diagnostic metric, rather than introduced as an additional design objective.

After solving \((\mathcal P_{\rm OA})\), the calibrated analyzer is denoted by
\begin{align}
\widehat{\mathcal Q}_{\rm OA}
    =
    \widehat{\alpha}_2
    \mathcal Q_{\rm OA}(\widehat{\boldsymbol\Omega}_2),
\label{eq:sim2_calibrated_oa}
\end{align}
and reused across channel realizations. The scalar \(\widehat{\alpha}_2\) is a fixed analyzer calibration gain obtained during offline fitting.

\vspace{-0.5cm}
\subsection{SIM$_3$: Residual MIMO Detection}
\label{subsec:sim3_module}
\label{subsec:det_sim3_model}

After SIM$_1$ and SIM$_2$, the receive waveform has been converted into the stacked virtual OFDM-analysis vector
\begin{align}
\mathbf y_b
=
\left[
\mathbf y_b^T[0],
\mathbf y_b^T[1],
\ldots,
\mathbf y_b^T[K-1]
\right]^T
\in\mathbb C^{KM},
\end{align}
where \(k\) is a logical OFDM-mode index. SIM$_3$ maps this \(KM\)-element interface to a \(KS\)-element stream-output interface. Similar to the virtual-subcarrier grouping used by the transmitter-side SIM precoder, SIM$_3$ is one unified block-structured transfer matrix whose input and output coordinates are ordered by virtual subcarrier. The overall \(KS\times KM\) detector therefore contains \(K\) diagonal \(S\times M\) detection blocks within a single SIM section, rather than \(K\) physically independent detector devices.

\subsubsection{Architecture and Operator Model}

The SIM$_3$ wave-domain operator \(\mathcal Q_{\rm DET}(\boldsymbol\Omega_3)\in\mathbb C^{KS\times KM}\) is written compactly as
\begin{align}
\mathcal Q_{\rm DET}(\boldsymbol\Omega_3)
=
\mathbf S_3
\boldsymbol{\Psi}_3^{L_3}
\mathbf W_3^{L_3}
\cdots
\boldsymbol{\Psi}_3^{1}
\mathbf W_3^{1},
\label{eq:qdet_full_sim3}
\end{align}
where the phase configuration is \(\boldsymbol\Omega_3=\{\boldsymbol{\Psi}_3^l\}_{l=1}^{L_3}\), \(\boldsymbol{\Psi}_{3}^{l}\) is the phase-shift matrix of the \(l\)-th layer, \(\mathbf W_3^{l}\) follows the common SIM coupling convention, and \(\mathbf S_3\in\mathbb C^{KS\times d_{3,L_3}}\) reads out the stream-output interface. 

To preserve the OFDM-mode grouping recovered by SIM$_2$, the effective detector is configured toward the block-diagonal form with \(K\) diagonal subblocks, each of size \(S\times M\),
\begin{align}
\mathcal Q_{\rm DET}
=
{\rm blkdiag}
\left(
\mathbf Q_{{\rm DET},0},
\mathbf Q_{{\rm DET},1},
\ldots,
\mathbf Q_{{\rm DET},K-1}
\right),
\label{eq:qdet_block}
\end{align}
where \(\mathbf Q_{{\rm DET},k}\in\mathbb C^{S\times M}\) detects the \(S\) streams on virtual subcarrier \(k\) from the corresponding \(M\)-element SIM$_2$ output group. The detector output is
\begin{align}
\hat{\mathbf s}_b
=
\mathcal Q_{\rm DET}(\boldsymbol\Omega_3)\mathbf y_b \in\mathbb C^{KS}.
\end{align}

\subsubsection{Configuration Target}

With SIM$_1$ and SIM$_2$ fixed, the remaining linear operation is residual MIMO detection over the virtual OFDM-analysis interface. In the following, \(\mathcal Q_{\rm CS}\) denotes the configured SIM$_1$ shortener in \eqref{eq:sim1_block_toeplitz}, while \(\widehat{\mathcal Q}_{\rm OA}\) denotes the calibrated SIM$_2$ analyzer in \eqref{eq:sim2_calibrated_oa}. The front-end equivalent channel before SIM$_3$ is
\begin{align}
\mathbf T_{12}
=
\widehat{\mathcal Q}_{\rm OA}
\mathcal Q_{\rm CS}
\bar{\mathbf H}_{\rm conv}
\mathbf Q_{\rm TX},
\label{eq:t12_frontend}
\end{align}
and is partitioned into \(\{\mathbf T_{12,k,q}\}_{k,q=0}^{K-1}\), where \(\mathbf T_{12,k,q}\in\mathbb C^{M\times S}\). The corresponding front-end noise covariance is
\begin{align}
\mathbf R_y
=
\sigma^2
\widehat{\mathcal Q}_{\rm OA}
\mathcal Q_{\rm CS}
\mathcal Q_{\rm CS}^{H}
\widehat{\mathcal Q}_{\rm OA}^{H}.
\label{eq:front_end_noise_cov}
\end{align}

Let \(\mathbf E_k^{y}\in\{0,1\}^{KM\times M}\) select the \(M\) SIM$_2$ output elements associated with virtual subcarrier \(k\). Then
\begin{align}
\mathbf R_{n,k}
=
(\mathbf E_k^{y})^T
\mathbf R_y
\mathbf E_k^{y}.
\label{eq:subcarrier_noise_cov}
\end{align}

For subcarrier \(k\), the SIM$_2$ output can be written as
\begin{align}
\mathbf y_b[k]
=
\mathbf T_{12,k,k}\mathbf s_b[k]
+
\sum_{\substack{q=0\\q\ne k}}^{K-1}
\mathbf T_{12,k,q}\mathbf s_b[q]
+
\mathbf n_y[k].
\end{align}

The residual interference-plus-noise covariance is
\begin{align}
\mathbf R_{\nu,k}
=
\sum_{\substack{q=0\\q\ne k}}^{K-1}
\mathbf T_{12,k,q}
\mathbf R_s[q]
\mathbf T_{12,k,q}^{H}
+
\mathbf R_{n,k},
\end{align}
where \(\mathbf R_s[q]\) is the covariance of \(\mathbf s_b[q]\). Previous-block leakage, if present in \eqref{eq:block_conv_with_ibi}, is included in this residual covariance. The colored-interference-aware MMSE target is
\begin{align}
\mathbf W_{\rm MMSE}[k]
=
\mathbf T_{12,k,k}^{H}
\left(
\mathbf T_{12,k,k}\mathbf T_{12,k,k}^{H}
+
\mathbf R_{\nu,k}
\right)^{-1}.
\end{align}

The diagonal detector block is
\begin{align}
\mathbf Q_{{\rm DET},k}(\boldsymbol\Omega_3)
=
(\mathbf E_k^{\rm out})^T
\mathcal Q_{\rm DET}(\boldsymbol\Omega_3)
\mathbf E_k^{\rm in},
\end{align}
where \(\mathbf E_k^{\rm in}\in\{0,1\}^{KM\times M}\) and \(\mathbf E_k^{\rm out}\in\{0,1\}^{KS\times S}\) select the \(k\)-th input and output interface groups, respectively.

The block-structured fitting problem is
\begin{subequations}
\begin{align}
(\mathcal P_{\rm DET}):~
\min_{\boldsymbol\Omega_3,\{\alpha_{3,k}\}}\quad&
\mathcal L_{\rm DET}\\
{\rm s.t.}\quad&
\theta_{3,m}^{l}\in\mathcal B_3,\quad \forall l,m,
\end{align}
\label{prob:det_fit_sec3}
\end{subequations}
\!\!\!where
\begin{align}
\mathcal L_{\rm DET}
=
\sum_{k=0}^{K-1}
\frac{
\left\|
\alpha_{3,k}
\mathbf Q_{{\rm DET},k}(\boldsymbol\Omega_3)
-
\mathbf W_{\rm MMSE}[k]
\right\|_F^2
}{
\left\|
\mathbf W_{\rm MMSE}[k]
\right\|_F^2
},
\end{align}
and the scalar \(\alpha_{3,k}\in\mathbb C\) is a detector-block calibration factor. 

After solving \((\mathcal P_{\rm DET})\), the calibrated detector is
\begin{equation}
    \widehat{\mathcal Q}_{\rm DET}
    =
    \widehat{\mathbf A}_3
    \mathcal Q_{\rm DET}(\widehat{\boldsymbol\Omega}_3),
\label{eq:sim3_calibrated_detector}
\end{equation}
where $\widehat{\mathbf A}_3
    =
    {\rm blkdiag}
    (
    \widehat{\alpha}_{3,0}\mathbf I_S,\ldots,
    \widehat{\alpha}_{3,K-1}\mathbf I_S
    ).$

\subsection{Cascaded End-to-End Receiver Matrix}
\label{sec:e2e_receiver_matrix}

This subsection collects the cascaded receiver matrix used for system-level diagnostics after SIM configuration. The normalized cross-subcarrier coupling of the SIM$_1$ and SIM$_2$ front-end in \eqref{eq:t12_frontend} is
\begin{align}
\rho_{\rm ICI}^{(1,2)}
=
\frac{
\sum_{k=0}^{K-1}
\sum_{\substack{q=0\\q\ne k}}^{K-1}
\left\|
\mathbf T_{12,k,q}
\right\|_F^2
}{
\sum_{k=0}^{K-1}
\sum_{q=0}^{K-1}
\left\|
\mathbf T_{12,k,q}
\right\|_F^2+\epsilon_0
},
\label{eq:rho_ici_12}
\end{align}
which measures the residual ICI before SIM$_3$. 

The complete wave-domain linear receiver output is
\begin{align}
\hat{\mathbf s}_b
&=
\widehat{\mathcal Q}_{\rm DET}
\widehat{\mathcal Q}_{\rm OA}
\mathcal Q_{\rm CS}
\bar{\mathbf r}_b \nonumber\\
&=
\mathbf T_3
\mathbf s_b
+
\tilde{\mathbf n}_b,
\label{eq:e2e_t3}
\end{align}
where \(\mathbf T_3=\widehat{\mathcal Q}_{\rm DET}\mathbf T_{12}\) and
\(
\tilde{\mathbf n}_b
=
\widehat{\mathcal Q}_{\rm DET}
\widehat{\mathcal Q}_{\rm OA}
\mathcal Q_{\rm CS}
\bar{\mathbf w}_b
\).
The matrix \(\mathbf T_3\) is partitioned into \(S\times S\) blocks \(\{\mathbf T_{3,k,q}\}_{k,q=0}^{K-1}\) to inspect the complete wave-domain linear receiver after SIM$_3$. If the previous-block term in \eqref{eq:block_conv_with_ibi} is nonzero, its contribution is treated as residual interference in the detector covariance.

\begin{remark}
The low-dimensional control branch provides coarse timing, residual CFO compensation, CSI acquisition, and SIM configuration. These operations only trigger and calibrate the receiver; the high-dimensional CP removal, DFT analysis, and linear detection remain in the wave domain. With residual timing uncertainty, SIM$_2$ may evaluate a small candidate set \(\mathcal D_{\tau}\):
\begin{align}
\mathbf y_b^{(\delta)}
=
\widehat{\mathcal Q}_{\rm OA}^{(\delta)}
\mathcal Q_{\rm CS}
\bar{\mathbf r}_b,
\qquad
\delta\in\mathcal D_{\tau}.
\label{eq:multi_hypothesis_oa}
\end{align}
Here \(\widehat{\mathcal Q}_{\rm OA}^{(\delta)}\) denotes the calibrated SIM$_2$ analyzer with the CP-window selection shifted by \(\delta\) samples.

The selected candidate minimizes a low-dimensional pilot residual,
\begin{align}
\mathcal M_{\rm pil}^{(\delta)}
=&
\sum_{k\in\mathcal K_{\rm p}}
\left\|
\widehat{\mathbf s}_{\rm p}^{(\delta)}[k]
-
\mathbf s_{\rm p}[k]
\right\|_2^2, \\
\hat{\delta}_b
&=
\arg\min_{\delta\in\mathcal D_{\tau}}
\mathcal M_{\rm pil}^{(\delta)},
\label{eq:timing_hypothesis_selection}
\end{align}
where \(\mathcal K_{\rm p}\) is a small pilot-subcarrier set, \(\mathbf s_{\rm p}[k]\) is the known pilot, and \(\widehat{\mathbf s}_{\rm p}^{(\delta)}[k]\) is the corresponding low-dimensional pilot estimate. The selected \(\mathbf y_b^{(\hat{\delta}_b)}\) is routed to SIM$_3$. As a system-level stress test for channel-delay uncertainty, we perturb the effective channel delay profile by adjacent-tap leakage:
\begin{align}
\widetilde{\mathbf H}[\ell]
&=
\left(1-\epsilon_{\tau}|\xi_{\ell}|\right)\mathbf H[\ell]
+
\epsilon_{\tau}\xi_{\ell-1}\mathbf H[\ell-1],
\label{eq:delay_profile_perturbation}
\end{align}
where \(\epsilon_{\tau}\) is the normalized perturbation level, \(\xi_{\ell}\in[-1,1]\), \(\xi_{-1}=0\), and \(\mathbf H[-1]=\mathbf 0\).
\end{remark}

\section{Solution Framework: Offline Compilation and Learning-Assisted Online Configuration}
\label{sec:solution_framework}

This section gives the configuration procedure for the module targets defined in Section~\ref{sec:receiver_modules}. The purpose is to explain how the three SIM sections are configured, rather than to introduce new receiver functions. SIM$_2$ is compiled offline because it depends only on the OFDM numerology, while SIM$_1$ and SIM$_3$ are configured online over the channel coherence time using compact architecture-aware controls.
\vspace{-0.5cm}
\subsection{Offline Compilation of SIM$_2$}
For fixed phase variables \(\boldsymbol\Omega_2\), the calibration gain in \(\mathcal P_{\rm OA}\) has the closed form
\begin{equation}
    \widehat {\alpha}_2
    =
    \frac{
    {\rm Tr}\left[
    \mathbf C_{\rm sum}^{H}(\boldsymbol\Omega_2)
    \mathbf C_{\rm sum}^{\star}
    \right]
    }{
    \left\|
    \mathbf C_{\rm sum}(\boldsymbol\Omega_2)
    \right\|_F^2
    }.
    \label{eq:alpha2_closed_form}
\end{equation}

The phase variables are updated by projected phase descent:
\begin{equation}
    \theta_{2,m}^{l,(t+1)}
    =
    \Pi_{\mathcal B_2}
    \left[
    \theta_{2,m}^{l,(t)}
    -
    \mu_2^{(t)}
    \frac{\partial \mathcal L_{\rm OA}}
    {\partial \theta_{2,m}^l}
    \right],
    \label{eq:oa_phase_update}
\end{equation}
where \(\Pi_{\mathcal B_2}[\cdot]\) projects onto the feasible phase codebook and \(\mu_2^{(t)}\) is the step size. The derivative is obtained by back-propagating through the layered SIM product. After each phase update, \(\widehat{\alpha}_2\) is recalculated using \eqref{eq:alpha2_closed_form}. The iterations stop when the operator fitting error no longer improves, while row-Gram leakage is monitored as a mode-orthogonality diagnostic. The output is the fixed calibrated analyzer \(\widehat{\mathcal Q}_{\rm OA}\) in \eqref{eq:sim2_calibrated_oa}.

\subsection{Learning-Assisted Configuration of SIM$_1$}
For SIM$_1$, learning is used as an amortized controller for compact channel-adaptive variables, while the receiver operator is still generated by the explicit model in \(\mathcal P_{\rm CS}\). Since directly optimizing \(\boldsymbol\Omega_U\), \(\boldsymbol\Omega_V\), and \(\{\boldsymbol\Lambda_p\}\) is difficult under the factorized tap structure, we use a target-and-materialization procedure: first generate a relaxed reference FIR shortener, and then project it onto the structured SIM$_1$ realization.

The online controller selects the two most influential weights for SIM$_1$ shortener generation,
\begin{align}
\boldsymbol\xi_1
=
\left[
\lambda_{\rm tail}^{\rm eff},
\lambda_{\rm reg}^{\rm eff}
\right]^T,
\label{eq:xi1_def}
\end{align}
where \(\lambda_{\rm tail}^{\rm eff}\) controls the out-of-CP tail emphasis in the weighted target generator and \(\lambda_{\rm reg}^{\rm eff}\) is the online instantiated value of the regularization coefficient \(\lambda_{\rm reg}\) in \((\mathcal P_{\rm CS})\). The controller selects interpretable shortener-generation parameters rather than directly outputting dense FIR coefficients.

Let \( \mathbf q_{\rm CS}={\rm vec} ( [\mathbf Q_0^{\rm CS},\mathbf Q_1^{\rm CS},\ldots,\mathbf Q_{P_1-1}^{\rm CS}] )\) collect all FIR coefficients. For a fixed channel, the effective-channel relation \eqref{eq:sim1_effectivechannel} is linear in \(\mathbf q_{\rm CS}\). By vectorizing this relation and stacking the selected effective-delay samples, we write the relaxed matching model as \(\mathbf A_{\rm eq}\mathbf q_{\rm CS}\approx\mathbf b_{\rm eq}\), where \(\mathbf A_{\rm eq}\) is determined by the current channel taps. The vector \(\mathbf b_{\rm eq}\) is not an all-zero target: its out-of-CP entries are set to zero, while its CP-compatible entries specify a normalized useful-response target so that the relaxed shortener preserves nonzero in-CP energy, consistent with \eqref{eq:cs_signal_constraint}. The reference shortener is generated by the weighted least-squares surrogate
\begin{align}
    \mathbf q_{{\rm CS},{\rm ref}}(\boldsymbol\xi_1)
    &=
    \operatorname*{arg}\operatorname*{min}_{\mathbf q_{\rm CS}} \nonumber \\
&\{
    \left\|
    \mathbf A_{\rm CS}(\boldsymbol\xi_1)\mathbf q_{\rm CS}
    -
    \mathbf b_{\rm CS}
    \right\|_2^2
    +
    \lambda_{\rm reg}^{\rm eff}
    \left\|
    \mathbf q_{\rm CS}
    \right\|_2^2\},
    \label{eq:cs_weighted_ls}
\end{align}
\begin{align}
\mathbf A_{\rm CS}(\boldsymbol\xi_1)
=
\mathbf W_{\rm tail}(\lambda_{\rm tail}^{\rm eff})\mathbf A_{\rm eq},
~
\mathbf b_{\rm CS}
=
\mathbf W_{\rm tail}(\lambda_{\rm tail}^{\rm eff})\mathbf b_{\rm eq},
\label{eq:acs_decomposition}
\end{align}
where the diagonal matrix \(\mathbf W_{\rm tail}(\lambda_{\rm tail}^{\rm eff})\) gives unit weight to CP-compatible equations and higher weight to out-of-CP equations. Thus, \(\lambda_{\rm tail}^{\rm eff}\) controls how aggressively the target generator suppresses the effective tail, while \(\lambda_{\rm reg}^{\rm eff}\) controls the FIR norm and the associated noise-enhancement proxy.

For fixed \(\boldsymbol\xi_1\), \eqref{eq:cs_weighted_ls} admits the normal-equation solution
\begin{equation}
    \mathbf q_{{\rm CS},{\rm ref}}(\boldsymbol\xi_1)
    =
    \left(
    \mathbf A_{\rm CS}^H\mathbf A_{\rm CS}
    +
    \lambda_{\rm reg}^{\rm eff}\mathbf I
    \right)^{-1}
    \mathbf A_{\rm CS}^H
    \mathbf b_{\rm CS},
    \label{eq:cs_normal_equation}
\end{equation}
where the vector \( \mathbf q_{{\rm CS},{\rm ref}}\) is reshaped into reference tap matrices \(\{\mathbf Q_{p,{\rm ref}}^{\rm CS}\}_{p=0}^{P_1-1}\). These matrices describe the desired channel-shortening FIR taps, but they are not yet constrained by the SIM$_1$ architecture. Hence, SIM$_1$ performs a hardware-aware structured projection: the reference taps are approximated by the beamspace-delay form. This materialization step, referred to as the SIM$_1$ structured projection, is formulated in \eqref{eq:sim1_structured_projection},
\begin{figure*}[h]
\begin{equation}
\{\boldsymbol\Omega_{U}^\star,\boldsymbol\Omega_{V}^\star,\{\boldsymbol\Lambda_p^\star\}_{p=0}^{P_1-1},\alpha_1^\star\}
=
\arg\min_{\boldsymbol\Omega_{U},\boldsymbol\Omega_{V},\{\boldsymbol\Lambda_p\},\alpha_1}
\frac{
\sum_{p=0}^{P_1-1}
\left\|
\alpha_1
\mathbf U(\boldsymbol\Omega_{U})
\boldsymbol\Lambda_p
\mathbf V(\boldsymbol\Omega_{V})
-
\mathbf Q_{p,{\rm ref}}^{\rm CS}
\right\|_F^2
}{
\sum_{p=0}^{P_1-1}
\left\|
\mathbf Q_{p,{\rm ref}}^{\rm CS}
\right\|_F^2
+\epsilon_1
},
\label{eq:sim1_structured_projection}
\end{equation}
\hrule 
\vspace{-0.5 cm}
\end{figure*}
where \(\alpha_1\in\mathbb C\) is a scalar calibration gain for SIM$_1$ tap materialization and \(\epsilon_1>0\) prevents numerical division by zero. After solving \eqref{eq:sim1_structured_projection}, the calibrated operator is denoted by
\begin{equation}
\widehat{\mathbf Q}_p^{\rm CS}=\widehat{\alpha}_1\mathbf U(\widehat{\boldsymbol\Omega}_{U})
\widehat{\boldsymbol\Lambda}_p
\mathbf V(\widehat{\boldsymbol\Omega}_{V}),
\end{equation}
where the taps \(\{\widehat{\mathbf Q}_p^{\rm CS}\}\) are then inserted into the block-Toeplitz form in \eqref{eq:sim1_block_toeplitz}, yielding the configured SIM$_1$ shortener \(\mathcal Q_{\rm CS}\).

The SIM$_1$ controller observes compact channel features
\begin{equation}
    \mathbf o_1
    =
    \phi_1(\widehat{\mathbf H}[0],\ldots,\widehat{\mathbf H}[L_h-1],{\rm SNR}),
    \label{eq:cs_state}
\end{equation}
where \(\phi_1(\cdot)\) may include tap energies, delay spread, initial CP energy ratio, singular values, and frequency-domain condition numbers. The controller outputs
\begin{equation}
    \mathbf a_1
    =
    \pi_{\vartheta_1}^{\rm CS}(\mathbf o_1) \in[-1,1]^2,
\end{equation}
where \(\pi_{\vartheta_1}^{\rm CS}\) is the architecture-aware SIM$_1$ controller parameterized by trainable weights \(\vartheta_1\), and \(\mathbf a_1=[a_{1,1},a_{1,2}]^T\) is its bounded action.

The bounded action is mapped to positive weights by logarithmic scaling:
\begin{align}
    \lambda_{\rm tail}^{\rm eff}
    &=
    10^{\ell_{\rm tail,min}
    +
    \frac{a_{1,1}+1}{2}
    (\ell_{\rm tail,max}-\ell_{\rm tail,min})},\\
    \lambda_{\rm reg}^{\rm eff}
    &=
    10^{\ell_{\rm reg,min}
    +
    \frac{a_{1,2}+1}{2}
    (\ell_{\rm reg,max}-\ell_{\rm reg,min})},
\end{align}
where \(\ell_{\rm tail,min}\), \(\ell_{\rm tail,max}\), \(\ell_{\rm reg,min}\), and \(\ell_{\rm reg,max}\) are preselected log-domain bounds for the two positive control weights.

The configured SIM$_1$ taps are then generated as
\begin{equation}
    \{\widehat{\mathbf Q}_p^{\rm CS}\}
    =
    \mathcal F_{\rm CS}
    \left(
    \widehat{\mathbf H};
    \lambda_{\rm tail}^{\rm eff},
    \lambda_{\rm reg}^{\rm eff}
    \right),
\label{eq:cs_control_operator_generation}
\end{equation}
where \(\mathcal F_{\rm CS}(\cdot)\) denotes the complete SIM$_1$ configuration map: it first generates the relaxed reference shortener by \eqref{eq:cs_weighted_ls}--\eqref{eq:cs_normal_equation} and then projects it onto the structured SIM$_1$ by \eqref{eq:sim1_structured_projection}.

The one-step reward is communication-oriented. Let \(R_{\rm ref}\) denote the sum spectral efficiency achieved by the SIM$_1$/SIM$_2$ front-end followed by the digital detector reference, and define \(\bar R_{\rm ref}=R_{\rm ref}/(KS)\). The reward used to select the compact control variables is
\begin{equation}
    \mathcal R_{\rm CS}
    =
    \bar R_{\rm ref}
    -
    \kappa_{\rm tail}\mathcal L_{\rm tail}
    -
    \kappa_{\rm ICI}\rho_{\rm ICI}^{(1,2)}
    -
    \kappa_n\log_{10}\left(1+\mathcal L_{\rm noise}^{(1)}\right),
    \label{eq:cs_reward}
\end{equation}
where \(\kappa_{\rm tail}\), \(\kappa_{\rm ICI}\), and \(\kappa_n\) are dimensionless reward-scaling constants. Since all penalty terms are normalized, the reward is used only to rank candidate SIM$_1$ configurations for the current channel state. The action is selected once per coherence block, so this is a contextual continuous-control problem rather than a long-horizon sequential decision problem.

\subsection{Learning-Assisted Configuration of SIM$_3$}
With SIM$_1$ and SIM$_2$ fixed, SIM$_3$ operates on the \(KM\)-dimensional virtual OFDM-analysis interface. Rather than outputting a dense \(KS\times KM\) detector, the controller selects loaded-MMSE parameters for each virtual-subcarrier block.

For subcarrier \(k\), define the loaded detector
\begin{equation}
    \mathbf W_k(\boldsymbol\xi_{3,k})
    =
    \mathbf T_{12,k,k}^H
    \left(
    \mathbf T_{12,k,k}\mathbf T_{12,k,k}^H
    +
    \beta_k\mathbf R_{\nu,k}
    +
    \eta_k c_k\mathbf I_{M}
    \right)^{-1},
    \label{eq:loaded_mmse_detector}
\end{equation}
where \(\boldsymbol\xi_{3,k}=[\beta_k,\eta_k]^T\) controls the interference-plus-noise loading and diagonal robustness loading, and
\begin{equation}
    c_k
    =
    \frac{1}{M}
    {\rm Tr}
    \left(
    \mathbf T_{12,k,k}\mathbf T_{12,k,k}^H
    +
    \mathbf R_{\nu,k}
    \right)
\end{equation}
normalizes the loading level.

The local detector state is
\begin{equation}
    \mathbf o_{3,k}
    =
    \phi_3(\mathbf T_{12,k,k},\mathbf R_{\nu,k},{\rm SNR}),
\end{equation}
where \(\phi_3(\cdot)\) is a deterministic feature map containing equivalent-channel and interference-plus-noise statistics. The controller outputs
\begin{equation}
    \mathbf a_{3,k}
    =
    \pi_{\vartheta_3}^{\rm DET}(\mathbf o_{3,k})
    \in[-1,1]^2,
\end{equation}
where \(\pi_{\vartheta_3}^{\rm DET}\) is the detector-loading controller parameterized by trainable weights \(\vartheta_3\). The bounded action is mapped to the positive loading factors \(\beta_k\) and \(\eta_k\) by logarithmic scaling over preselected bounds. The reward is
\begin{equation}
    r_{3,k}
    =
    R_k(\mathbf W_k)
    -
    \kappa_{\rm det}
    \left\|
    \mathbf W_k\mathbf T_{12,k,k}
    -
    \mathbf I_S
    \right\|_F^2,
    \label{eq:det_reward}
\end{equation}
where \(R_k(\mathbf W_k)\) is the spectral-efficiency contribution of subcarrier \(k\) under detector \(\mathbf W_k\), and \(\kappa_{\rm det}\ge0\) is a dimensionless penalty weight.

The resulting \(\mathbf W_k(\boldsymbol\xi_{3,k})\) is fitted by the \(k\)-th diagonal virtual-subcarrier block of SIM$_3$:
\begin{align}
    \min_{\boldsymbol\Omega_3,\{\alpha_{3,k}\}}
    \sum_{k=0}^{K-1}
    \frac{
    \left\|
    \alpha_{3,k}
    \mathbf Q_{{\rm DET},k}(\boldsymbol\Omega_3)
    -
    \mathbf W_k(\boldsymbol\xi_{3,k})
    \right\|_F^2
    }{
    \left\|
    \mathbf W_k(\boldsymbol\xi_{3,k})
    \right\|_F^2
    }.
    \label{eq:det_loaded_fitting}
\end{align}

For fixed SIM phases, \(\alpha_{3,k}\) is updated based on the least-squares criterion
\begin{equation}
    \widehat{\alpha}_{3,k}
    =
    \frac{
    {\rm Tr}
    \left[
    \mathbf Q_{{\rm DET},k}^{H}(\boldsymbol\Omega_3)
    \mathbf W_k(\boldsymbol\xi_{3,k})
    \right]
    }{
    \left\|
    \mathbf Q_{{\rm DET},k}(\boldsymbol\Omega_3)
    \right\|_F^2
    }.
    \label{eq:alpha3_closed_form}
\end{equation}

The SIM$_3$ phases are updated by projected phase descent:
\begin{equation}
    \theta_{3,m}^{l,(t+1)}
    =
    \Pi_{\mathcal B_3}
    \left[
    \theta_{3,m}^{l,(t)}
    -
    \mu_3^{(t)}
    \frac{\partial \mathcal L_{\rm DET}}
    {\partial \theta_{3,m}^l}
    \right],
    \label{eq:det_phase_update}
\end{equation}
where \(\Pi_{\mathcal B_3}[\cdot]\) projects onto the feasible SIM$_3$ phase codebook, \(\mu_3^{(t)}\) is the step size, and the derivative is back-propagated through the unified layered SIM$_3$ transfer matrix. A calibrated block-structured detector codebook may also be used when faster reconfiguration is required.

\subsection{Cascaded Online Configuration Protocol}

After the three SIM sections have been specified, the cascaded receiver is configured by combining offline compilation and coherence-block online adaptation. The learning-assisted controllers are trained offline over a channel distribution and deployed as low-latency online controllers. Let \(\mathcal D_{\rm tr}\) denote training channels from the target propagation environment. The offline training procedure is:
\begin{enumerate}
    \item Fit SIM$_2$ to obtain \(\widehat{\mathcal Q}_{\rm OA}\).
    \item Compute expert controls for SIM$_1$ over \(\boldsymbol\xi_1\) and warm-start \(\pi_{\vartheta_1}^{\rm CS}\).
    \item Refine \(\pi_{\vartheta_1}^{\rm CS}\) using reward \eqref{eq:cs_reward}.
    \item Compute detector loading controls \(\boldsymbol\xi_{3,k}\) to warm-start \(\pi_{\vartheta_3}^{\rm DET}\).
    \item Refine \(\pi_{\vartheta_3}^{\rm DET}\) using reward \eqref{eq:det_reward}, and fit or tabulate the block-structured SIM$_3$ configurations.
\end{enumerate}

During online configuration in Algorithm~\ref{alg:solution_online}, the control branch estimates \(\widehat{\mathbf H}\), block timing, and residual CFO, while the main receive signal processing remains wave-domain.

\begin{algorithm}[t]
\caption{Online Configuration of the Cascaded Wave-Domain Receiver}
\label{alg:solution_online}
\DontPrintSemicolon
\LinesNumbered
\KwIn{Estimated channel taps \(\{\widehat{\mathbf H}[\ell]\}\), SNR estimate, timing estimate \(\hat n_b\), fixed \(\widehat{\mathcal Q}_{\rm OA}\), trained controllers \(\pi_{\vartheta_1}^{\rm CS}\) and \(\pi_{\vartheta_3}^{\rm DET}\).}
\KwOut{Configured functional sections SIM$_1$, SIM$_2$, and SIM$_3$ for the current coherence block.}
Use \(\pi_{\vartheta_1}^{\rm CS}\) to select \(\boldsymbol\xi_1\) from \(\{\widehat{\mathbf H}[\ell]\}\) and the SNR estimate.\;
Generate \(\{\mathbf Q_{p,{\rm ref}}^{\rm CS}\}\) by \eqref{eq:cs_weighted_ls}--\eqref{eq:cs_normal_equation}, project it by \eqref{eq:sim1_structured_projection}, and assemble \(\mathcal Q_{\rm CS}\).\;
Trigger the fixed SIM$_2$ analyzer \(\widehat{\mathcal Q}_{\rm OA}\) using \(\hat n_b\), and obtain \(\{\mathbf T_{12,k,k},\mathbf R_{\nu,k}\}\).\;
For each virtual subcarrier \(k\), use \(\pi_{\vartheta_3}^{\rm DET}\) to generate \(\mathbf W_k(\boldsymbol\xi_{3,k})\) by \eqref{eq:loaded_mmse_detector}.\;
Embed \(\{\mathbf W_k\}_{k=0}^{K-1}\) into the block-structured SIM$_3$ target and realize \(\widehat{\mathcal Q}_{\rm DET}\) by \eqref{eq:det_loaded_fitting}.\;
Receive the data block through SIM$_1$/SIM$_2$/SIM$_3$; perform symbol slicing and decoding digitally.\;
\end{algorithm}
\vspace{-0.5cm}

\section{Results and Analysis}
\label{sec:results}

\subsection{Simulation Setup}
\label{subsec:sim_setup}

We consider a MIMO-OFDM receiver with \(K\!=\!16\) subcarriers, \(S=4\) streams, \(N_t=6\) transmit ports, \(M_r=6\) receive ports, and \(M=6\) shortened wave-domain outputs after SIM$_1$. The simulations are conducted in a normalized discrete-time baseband model with carrier frequency \(f_c=28\) GHz and bandwidth \(B=100\) MHz, corresponding to sampling interval \(T_s=10\) ns. Unless otherwise stated, \(N_{\rm CP}=4\), so \(T=20\). The effective stream-to-receive channel is modeled as an \(L_h=8\)-tap Rayleigh fading channel with an exponential power-delay profile and normalized average channel energy. The default setting is intentionally CP-insufficient because \(L_h-1=7>N_{\rm CP}=4\). A CP-sufficient case with \(N_{\rm CP}=8\) is also evaluated later. The SNR is swept from \(0\) to \(30\) dB in \(5\)-dB steps, and the link-level curves are averaged over \(70\) independent channel realizations and \(120\) quadrature phase shift keying (QPSK) OFDM blocks per SNR point. 

The hardware parameters follow Table~\ref{tab:sim_hardware_dimensions}. For SIM$_1$, the delay order is \(P_1=4\), and the compact beamspace dimension is set to \(R_{\rm c}=M_r=M=6\), while the internal programmable mixing layers contain \(L_V=L_U=3\) metasurface layers with \(d_{V,l}=d_{U,l}=96\) meta-atoms per layer. For SIM$_2$, the raw CP-extended block entering the OFDM analyzer has dimension \(TM=120\). The useful-window memory implements the CP-window selection with dimension \(KM=96\). The summation core uses \(L_2=6\) metasurface layers with \(d_{2,l}=240\) meta-atoms per layer. The input and output signal dimensions of SIM$_3$ are \(KM=96\) and \(KS=64\), respectively. The detector core uses \(L_3=4\) metasurface layers with \(d_{3,l}=48\) meta-atoms per layer, and its fitting is evaluated over $100$ channel realizations. For SIM$_1$, the deterministic controller grid is \(\lambda_{\rm tail}^{\rm eff}\in\{4,8,15,30,60\}\) and \(\lambda_{\rm reg}^{\rm eff}\in\{10^{-4},3\times10^{-4},10^{-3},3\times10^{-3},10^{-2}\}\).
\vspace{-0.5cm}
\subsection{Convergence Analysis of the Configuration Framework}

\begin{figure}[t]
    \centering
    \subfigure[\label{fig:convera}]{\includegraphics[width=0.23\textwidth]{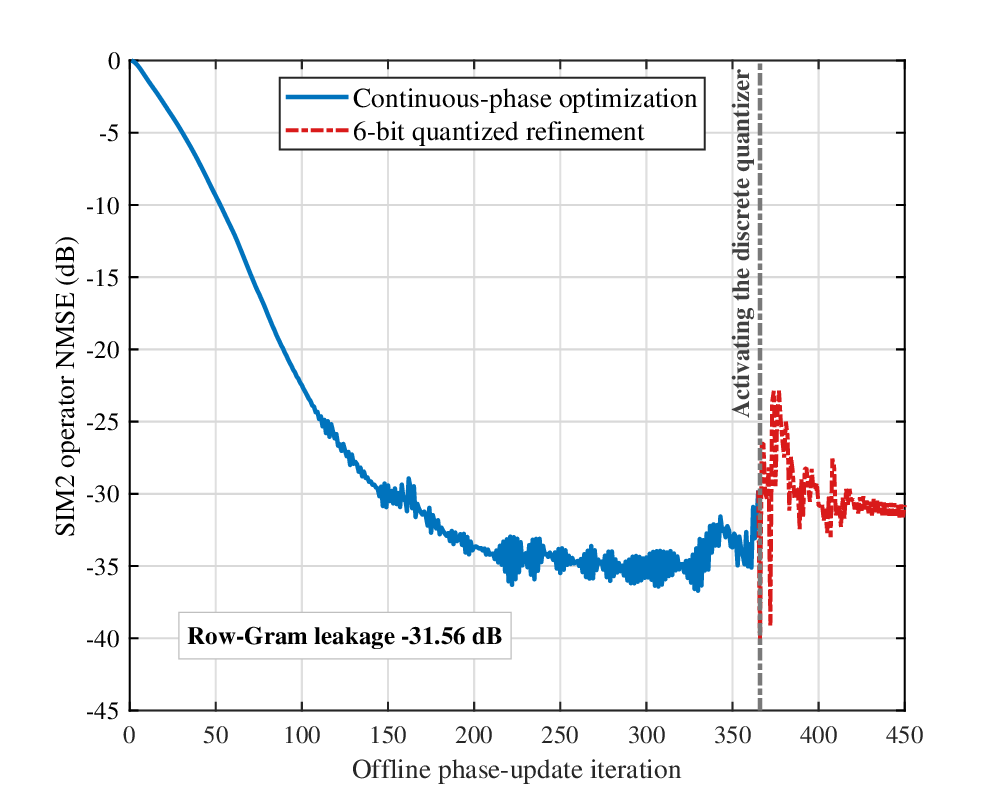}}
\hspace{-0.5 cm}
    \subfigure[\label{fig:converb}]{\includegraphics[width=0.23\textwidth]{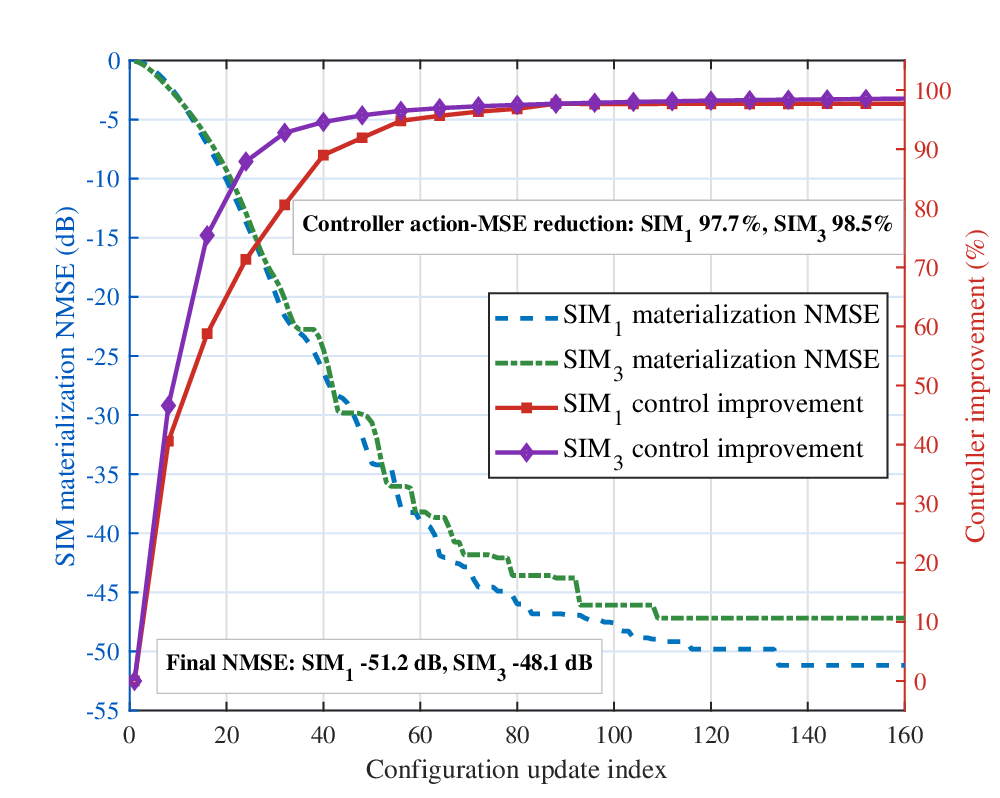}}
    \vspace{-0.2 cm}
\caption{Convergence behavior of the proposed wave-domain linear receiver: (a) Offline compilation of the SIM$_2$ OFDM analyzer; (b) Architecture-aware online configuration of the channel-adaptive SIM$_1$ and SIM$_3$ sections.}
    \label{fig:conver}
\vspace{-0.5 cm}
\end{figure}

Fig.~\ref{fig:convera} shows the offline compilation of SIM$_2$ for the fixed CP-windowed OFDM-analysis operator. The blue curve gives the operator NMSE with continuous phase shifts, while the red curve shows the result after activating the discrete 6-bit phase constraint. The operator NMSE rapidly decreases from the random initialization and remains in a low-error regime after the 6-bit phase constraint is activated. The final row-Gram leakage is \(-31.56\)~dB, confirming that the fitted analysis branches remain close to orthogonal. Since SIM$_2$ depends only on the OFDM numerology, this convergence behavior supports the intended offline compilation strategy: once calibrated, the same OFDM analyzer can be reused across channel realizations without online re-optimization.

Fig.~\ref{fig:converb} reports the online configuration behavior of SIM$_1$ and SIM$_3$. The left axis reports the materialization NMSE of the corresponding SIM operators, while the right axis reports the improvement of the architecture-aware learning controller, measured by the reduction of the validation action-MSE relative to the initial controller. The materialized channel-shortening and detector cores converge to about \(-51.2\)~dB and \(-48.1\)~dB NMSE, respectively, which verifies that the target operators generated by the model-based design can be accurately projected onto the phase-only SIM cores. Meanwhile, the learning-assisted controller reduces the validation action-MSE by \(97.7\%\) for SIM$_1$ and \(98.5\%\) for SIM$_3$. This result indicates that learning is used as an amortized controller for low-dimensional configuration variables, rather than as a black-box replacement of the receiver operator.

\vspace{-0.5 cm}
\subsection{Functional Validation of Wave-Domain Receiver}
\label{subsec:functional_validation}

In the functional validation, we first visualize SIM$_2$ because it provides the OFDM-analysis interface that directly replaces digital CP removal and DFT. We then show how SIM$_1$ prepares a CP-compatible waveform for this analyzer, followed by SIM$_3$ residual detection. 

\begin{figure}[t]
    \centering
    \subfigure[\label{fig:sim2a}]{\includegraphics[width=0.23\textwidth]{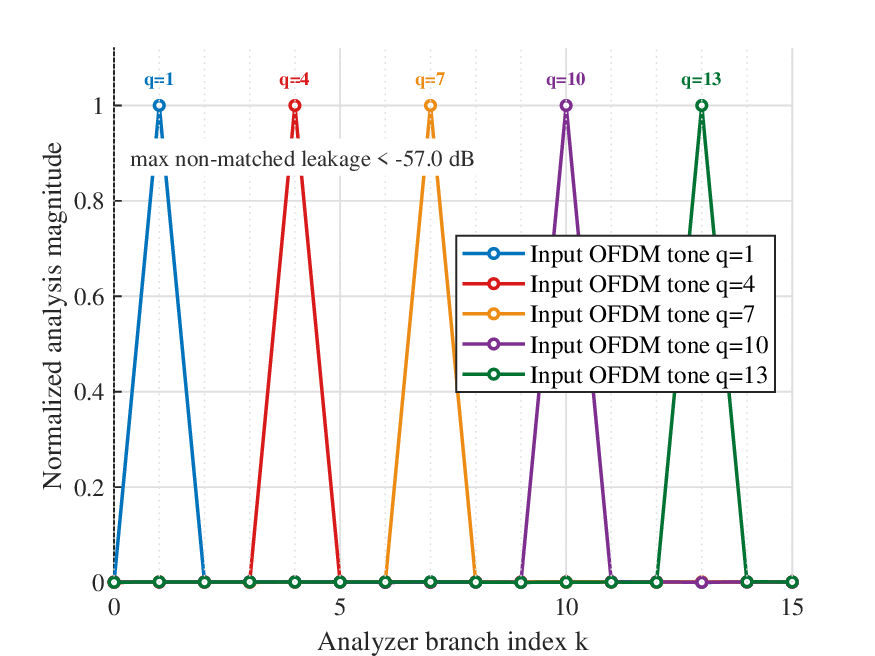}}
\hspace{-0.5 cm}
    \subfigure[\label{fig:sim2b}]{\includegraphics[width=0.23\textwidth]{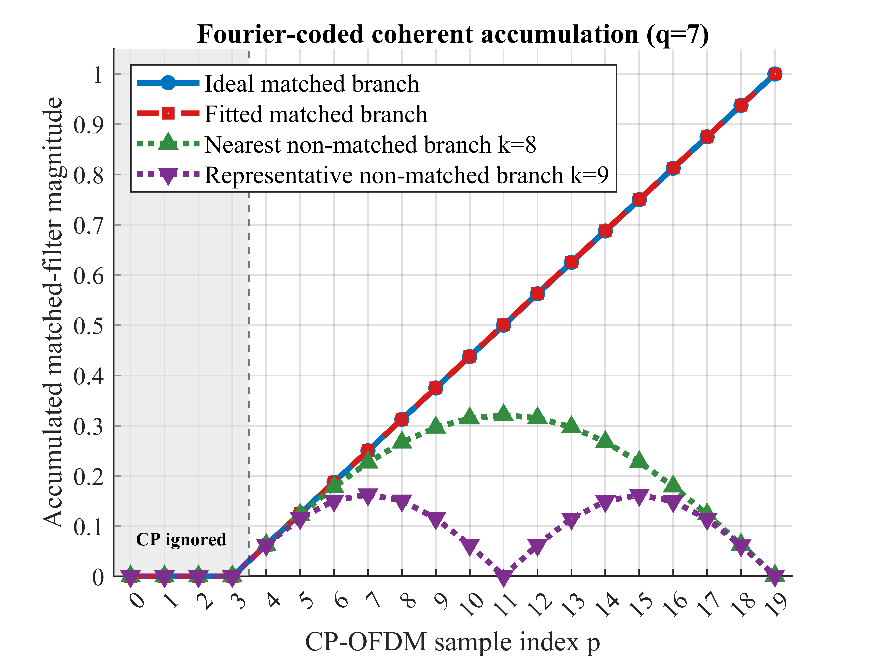}}
    \vspace{-0.2 cm}
\caption{Validation of the proposed SIM$_2$ wave-domain OFDM analyzer: (a) tone-to-branch selectivity of the fitted SIM$_2$ when representative CP-OFDM tones impinge on the receiver; (b) Fourier-coded coherent accumulation for an incident tone \(q=7\).}
    \label{fig:sim2_validation}
\vspace{-0.5 cm}
\end{figure}

Fig.~\ref{fig:sim2_validation} validates SIM$_2$ as a wave-domain OFDM analyzer. For each representative incident tone \(q\), Fig.~\ref{fig:sim2a} shows a dominant response at the matched branch \(k=q\), with strongly suppressed non-matched branches. This tone-to-branch selectivity is the wave-domain counterpart of subcarrier orthogonality in a digital DFT. Fig.~\ref{fig:sim2b} illustrates the corresponding accumulation for \(q=7\): the CP samples are assigned zero weight, the matched Fourier-coded branch adds coherently, and non-matched branches destructively cancel over the DFT window. Therefore, SIM$_2$ does not emulate CP removal and DFT as two separated digital operations; instead, it physically materializes their composite linear action as cyclic-windowed Fourier matched filtering in the wave domain.

\begin{figure}[t]
	\centerline{\includegraphics[width=0.45\textwidth]{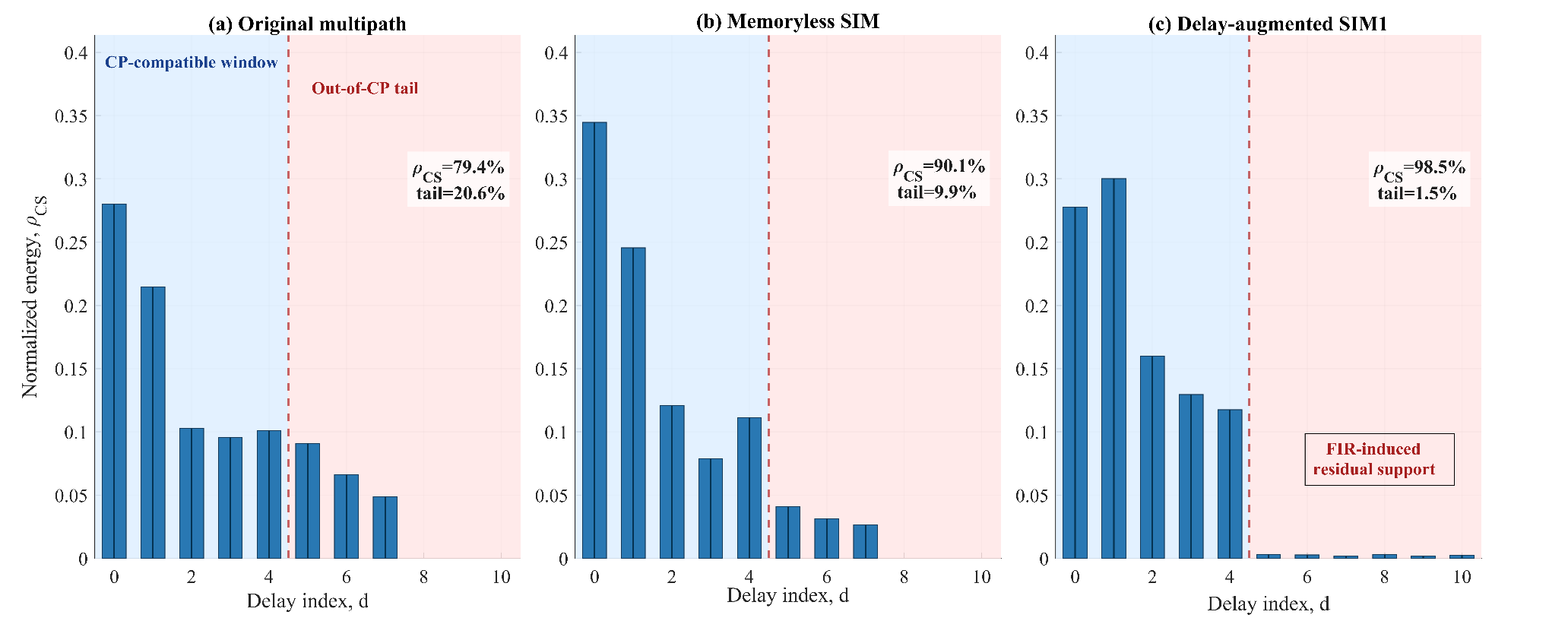}}
	\caption{The delay-domain energy profiles before OFDM analysis are compared for (a) the original multipath channel, (b) a memoryless SIM front-end, and (c) the proposed delay-augmented SIM$_1$. The blue region denotes the CP-compatible window, while the red region denotes the out-of-CP tail.}
	\label{fig:sim1_validation}
\vspace{-0.5 cm}
\end{figure}

Fig.~\ref{fig:sim1_validation} compares the normalized delay-energy profiles before OFDM analysis. The blue shaded region denotes the CP-compatible delay window, while the red shaded region denotes the out-of-CP tail. For fair comparison, all curves are plotted over the common effective support \(d=0,\ldots,L_h+P_1-2\). The original and memoryless cases are zero-padded beyond their native support. The memoryless SIM changes the spatial projection and slightly improves the in-CP energy ratio, but it cannot explicitly suppress the temporal tail because it has no access to delayed samples. In contrast, SIM$_1$ uses delay-augmented FIR processing to concentrate the effective channel energy inside the CP-compatible region and suppress the out-of-CP tail. The in-CP energy ratio increases from \(79.4\%\) for the original channel to \(90.1\%\) after memoryless SIM and to \(98.5\%\) after SIM$_1$, which verifies channel shortening in an energy-concentration sense before OFDM analysis.

\begin{figure}[t]
    \centering
    \subfigure[\label{fig:sim3a}]{\includegraphics[width=0.195\textwidth]{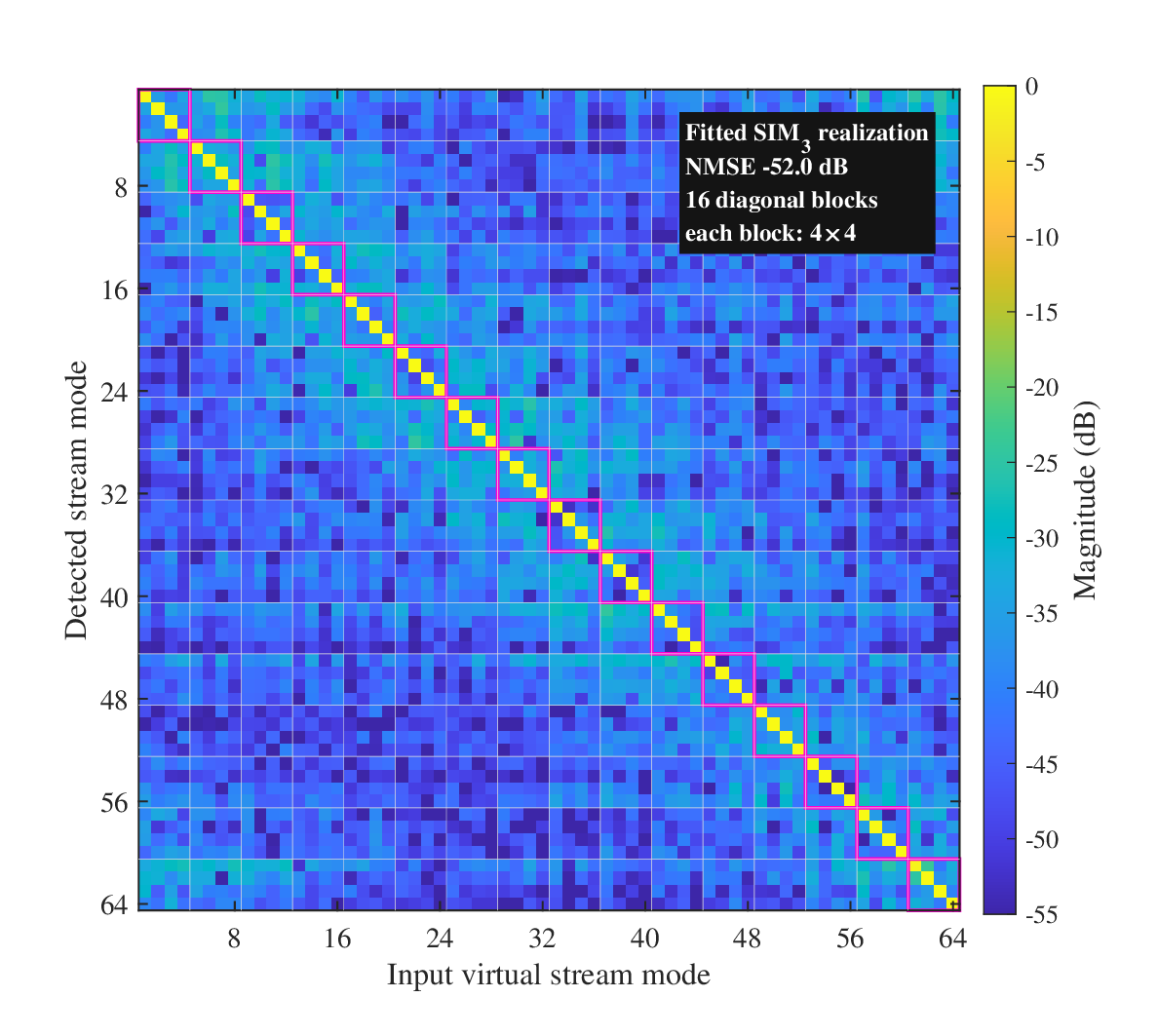}}
\hspace{-0.5 cm}
    \subfigure[\label{fig:sim3b}]{\includegraphics[width=0.23\textwidth]{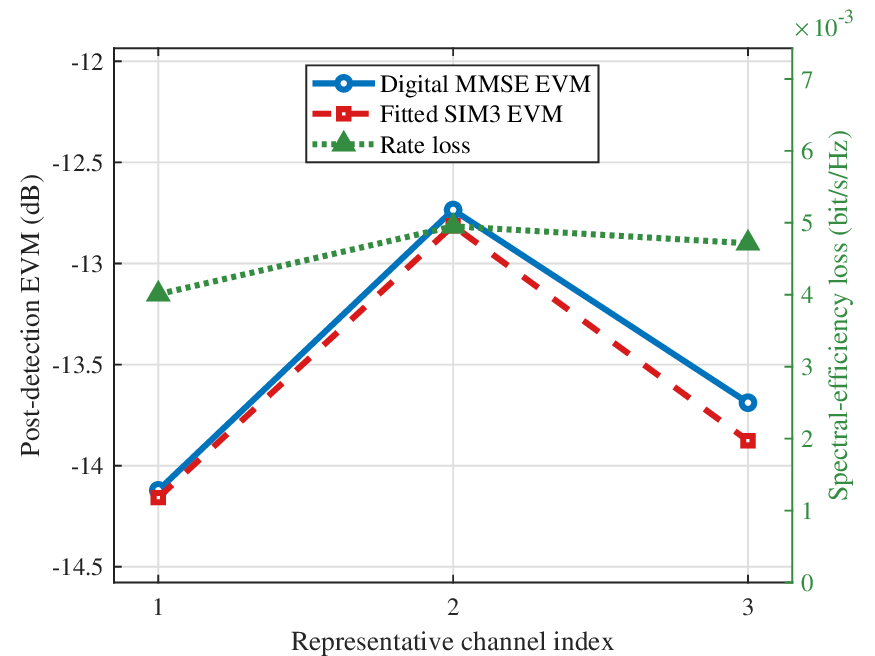}}
    \vspace{-0.2 cm}
\caption{Validation of SIM$_3$ virtual-subcarrier residual MIMO detection: (a) element-wise post-detection stream coupling matrix \(\mathbf T_3\) after the fitted SIM$_3$; (b) detection-output consistency between the digital MMSE reference and the fitted SIM$_3$.}
    \label{fig:sim3_validation}
\vspace{-0.5 cm}
\end{figure}

Fig.~\ref{fig:sim3_validation} validates SIM$_3$ over the virtual OFDM-analysis interface produced by SIM$_2$. Fig.~\ref{fig:sim3a} visualizes \(\mathbf T_3\) element by element. The highlighted \(S\times S\) diagonal blocks correspond to the \(K\) virtual subcarriers, while the weak off-block entries indicate suppressed cross-subcarrier leakage. Inside each highlighted block, the dominant element-wise diagonal entries show that the residual MIMO mixture on that virtual subcarrier has been converted into separated stream-output modes. Fig.~\ref{fig:sim3b} compares the digital MMSE reference and the fitted SIM$_3$ using error-vector magnitude (EVM) in $3$ representative examples, which measures the residual symbol distortion after detection and normalization. The fitted SIM$_3$ achieves a mean detector-fitting NMSE of \(-51.95\) dB and nearly overlaps with the digital MMSE reference, with \(4.6\times 10^{-3}\) bit/s/Hz effective-spectral-efficiency loss. This confirms that the final linear detector can also be moved to the wave-domain linear receive path with negligible replacement loss.
\vspace{-0.5 cm}
\subsection{End-to-End Wave-Domain Linear Reception Performance}
\label{subsec:performance}

We next evaluate the CP-insufficient setting with \(N_{\rm CP}=4\) and \(L_h=8\). The compared receivers are chosen to isolate the role of each functional section: 1) conventional digital DFT+MMSE receiver; 2) memoryless RX-SIM combiner with digital OFDM, which has no temporal memory; 3) SIM$_2$+SIM$_3$ without SIM$_1$, which removes the channel-shortening section; 4) SIM$_1$+SIM$_2$ without any digital or SIM-based detector, which serves as a negative control; 5) SIM$_1$+digital DFT+MMSE, which replaces SIM$_2$ by digital CP removal and DFT; 6) SIM$_1$+SIM$_2$+digital MMSE, which replaces SIM$_3$ by digital linear detection.

\begin{figure}[t]
	\centerline{\includegraphics[width=0.38\textwidth]{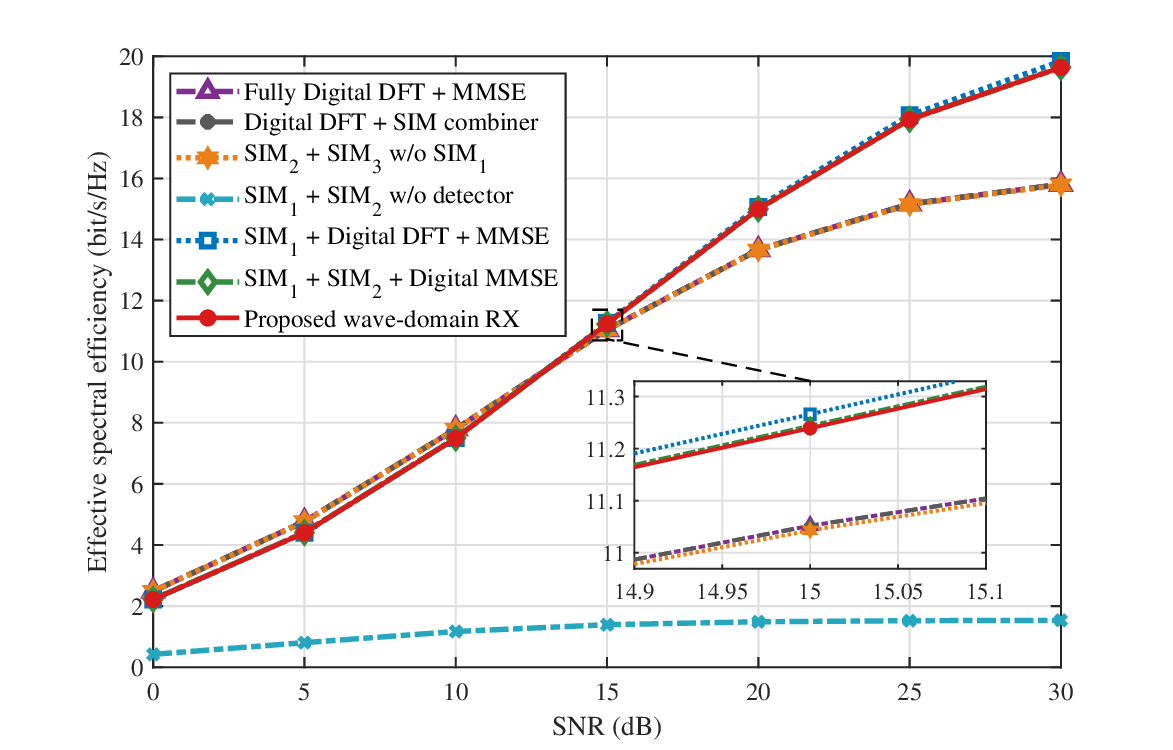}}
	\caption{Effective spectral efficiency under the CP-insufficient setting with \(N_{\rm CP}=4\) and \(L_h=8\). The inset magnifies the curves around \(15\) dB, excluding only the no-detector negative-control curve.}
	\label{fig:sr_cp_insufficient}
\vspace{-0.5 cm}
\end{figure}

Fig.~\ref{fig:sr_cp_insufficient} shows the effective spectral efficiency. The digital DFT+MMSE, memoryless RX-SIM, and SIM$_2$+SIM$_3$ without SIM$_1$ curves remain close at medium and high SNRs, indicating that post-OFDM processing and residual detection cannot remove the out-of-CP leakage created before OFDM analysis. Once SIM$_1$ is included, the high-SNR spectral efficiency increases substantially, which agrees with the delay-domain mechanism in Fig.~\ref{fig:sim1_validation}. The proposed wave-domain linear receiver closely follows the SIM$_1$+digital DFT+MMSE and SIM$_1$+SIM$_2$+digital MMSE references. Hence, the gain is mainly due to wave-domain pre-DFT channel-memory processing by SIM$_1$, while SIM$_2$ and SIM$_3$ introduce only a small replacement penalty when substituting digital CP/DFT analysis and MMSE combining.

\begin{figure}[t]
	\centerline{\includegraphics[width=0.38\textwidth]{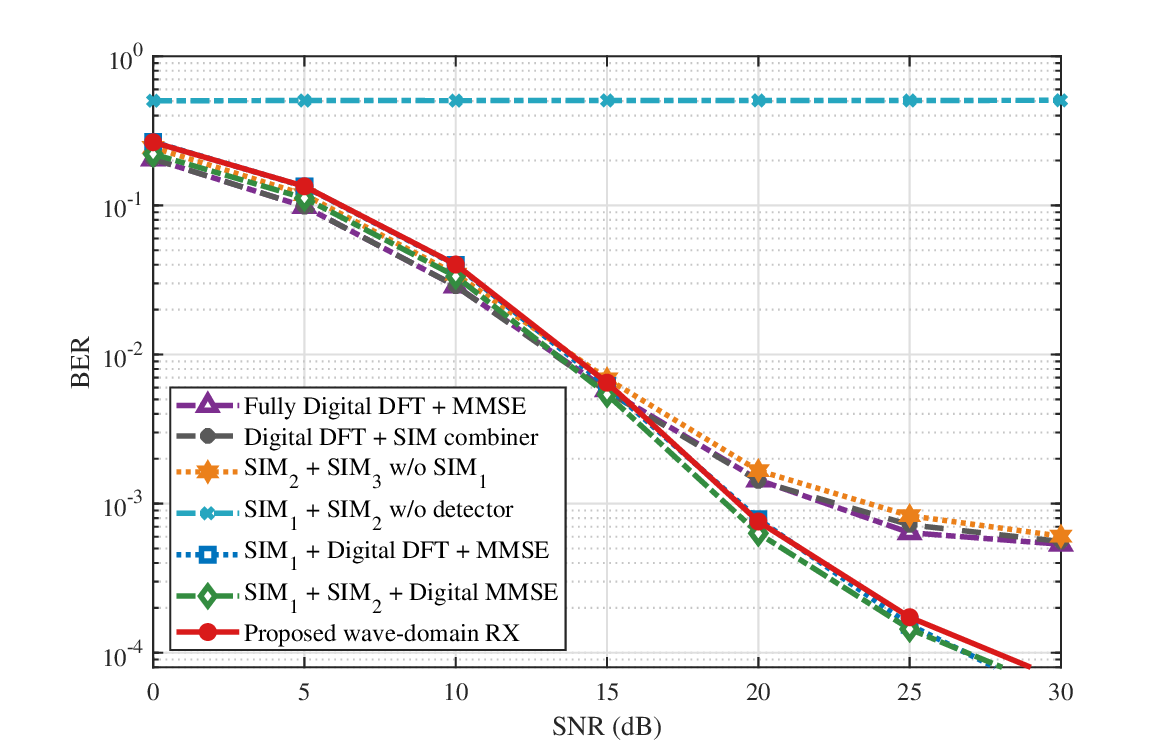}}
    \caption{BER performance under the CP-insufficient setting with \(N_{\rm CP}\!=\!4\) and \(L_h\!=\!8\), including functional-section ablation baselines.}
	\label{fig:ber_cp_insufficient}
\vspace{-0.5 cm}
\end{figure}

Fig.~\ref{fig:ber_cp_insufficient} reports the BER under the same CP-insufficient setting. The SIM$_1$+SIM$_2$ without detector curve acts as a negative control: channel shortening and OFDM analysis alone cannot separate the residual \(S\)-stream MIMO mixture. Baselines without SIM$_1$ exhibit a high-SNR error floor because residual out-of-CP leakage cannot be removed by a subcarrier-wise detector after the DFT stage. In contrast, SIM$_1$-assisted receivers continue to improve with SNR, showing that the energy-concentrated effective channel is more compatible with CP-OFDM analysis. The proposed wave-domain linear receiver remains close to the corresponding digital-detector and digital-OFDM references, confirming that SIM$_2$ and SIM$_3$ preserve the link-level reliability of their digital counterparts. In Table~\ref{tab:cp_regime_comparison}, \(R_{\rm SE}\) denotes the effective spectral efficiency.

\begin{table*}[t]
\centering
\caption{Performance Comparison under CP-Insufficient and CP-Sufficient Regimes}
\label{tab:cp_regime_comparison}
\begin{tabular}{llccccc}
\hline
& Receiver
& \(R_{\rm SE}\) at 20 dB
& BER at 20 dB
& \(R_{\rm SE}\) at 30 dB
& BER at 30 dB
& \(\Delta R_{\rm SE}\) at 30 dB \\
\hline
\multirow{4}{*}{Insufficient CP}
& Fully digital DFT+MMSE
& 13.6129 & \(1.44\times10^{-3}\)
& 15.7474 & \(6.71\times10^{-4}\)
& 0.00\% \\
& SIM$_1$ + digital DFT + MMSE
& 14.9478 & \(7.65\times10^{-4}\)
& 19.7064 & \(9.02\times10^{-5}\)
& \(+25.14\%\) \\
& SIM$_1$+SIM$_2$ + digital MMSE
& 14.8728 & \(8.36\times10^{-4}\)
& 19.4882 & \(9.58\times10^{-5}\)
& \(+23.75\%\) \\
& Proposed wave-domain RX
& 14.8682 & \(1.00\times10^{-3}\)
& 19.4836 & \(1.15\times10^{-4}\)
& \(+23.73\%\) \\
\hline
\multirow{5}{*}{Sufficient CP}
& Fully digital DFT+MMSE
& 14.6053 & \(1.21\times10^{-4}\)
& 23.2739 & 0 errors observed
& 0.00\% \\
& SIM$_1$ + digital DFT + MMSE
& 14.6054 & \(1.15\times10^{-4}\)
& 23.2740 & 0 errors observed
& \(+0.00\%\) \\
& SIM$_1$+SIM$_2$ + digital MMSE
& 14.5877 & \(1.29\times10^{-4}\)
& 23.0963 & 0 errors observed
& \(-0.76\%\) \\
& Proposed wave-domain RX
& 14.5839 & \(1.35\times10^{-4}\)
& 23.0925 & 0 errors observed
& \(-0.78\%\) \\
& Full-block LMMSE w/ CP obs.
& 15.8462 & \(1.86\times10^{-6}\)
& 24.5837 & 0 errors observed
& \(+5.63\%\) \\
\hline
\end{tabular}
\begin{flushleft}
\footnotesize
Notes: The entries labeled “0 errors observed” indicate that no bit errors were observed; they are not reported as strict BER upper bounds.
\end{flushleft}
\vspace{-0.6 cm}
\end{table*}

Table~\ref{tab:cp_regime_comparison} compares CP-insufficient and CP-sufficient regimes. Under insufficient CP, the conventional digital DFT+MMSE receiver is limited by out-of-CP leakage, while the proposed wave-domain linear receiver achieves a \(23.73\%\) effective-spectral-efficiency gain at 30 dB. This gain should therefore be attributed to the pre-DFT channel-memory processing of SIM$_1$, rather than to a stronger detector assumption. Under sufficient CP, the standard circular-convolution condition already holds, and the proposed receiver is intended to be a near-lossless functional replacement rather than a source of artificial gain. The effective spectral efficiency \(R_{\rm SE}=23.0925\) bit/s/Hz is obtained at 30 dB, only \(0.78\%\) below digital DFT+MMSE. The full-block LMMSE with CP observations is included as a stronger linear benchmark because it exploits CP samples discarded by the conventional DFT receiver.
\vspace{-0.5 cm}
\subsection{Robustness to Timing and Delay-Profile Perturbations}

The preceding sections assume that the low-dimensional control branch provides the timing trigger and that the channel-delay profile used for configuration matches the actual effective channel. This subsection examines how sensitive the proposed receiver is when these assumptions are imperfect. The purpose is not to replace a full synchronization receiver, but to verify that the wave-domain OFDM analyzer can tolerate small residual timing errors through candidate-window readout and that the cascaded receiver remains stable when neighboring channel taps are slightly mixed.

\begin{figure*}[t]
    \centering
    \subfigure[\label{fig:misa}]{\includegraphics[width=0.3\textwidth]{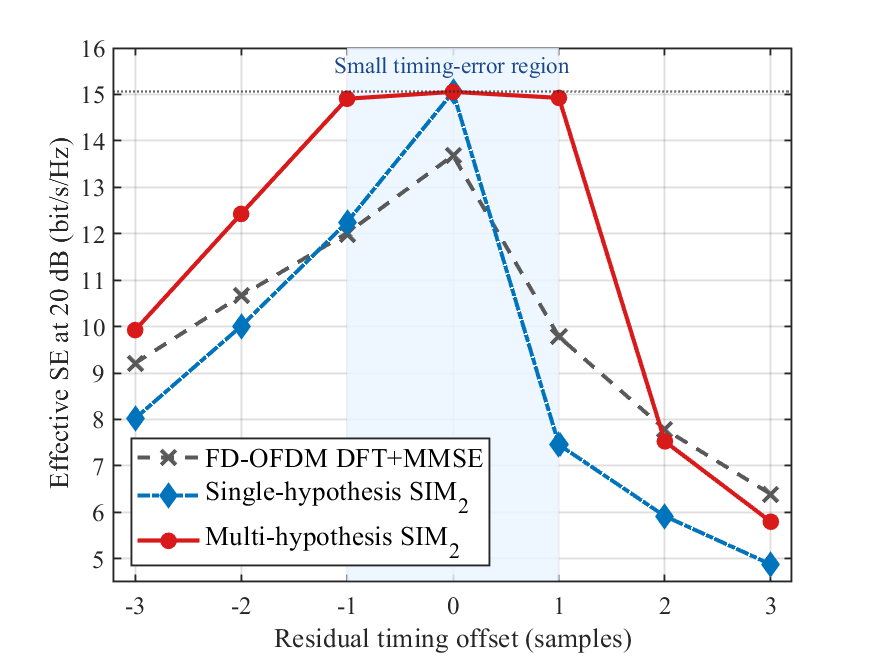}}
\hspace{-0.5 cm}
    \subfigure[\label{fig:misb}]{\includegraphics[width=0.3\textwidth]{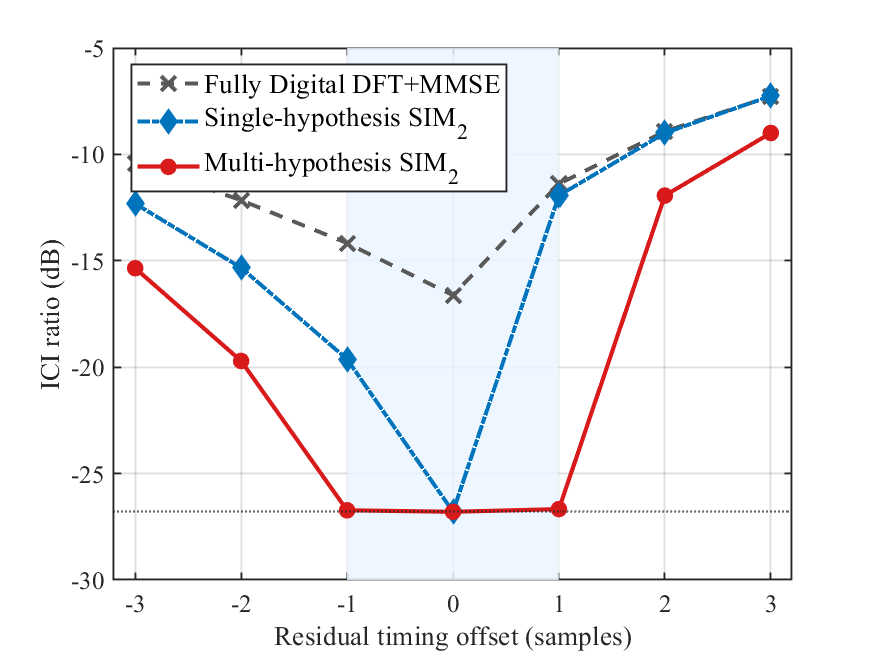}}
\hspace{-0.5 cm}
    \subfigure[\label{fig:misc}]{\includegraphics[width=0.3\textwidth]{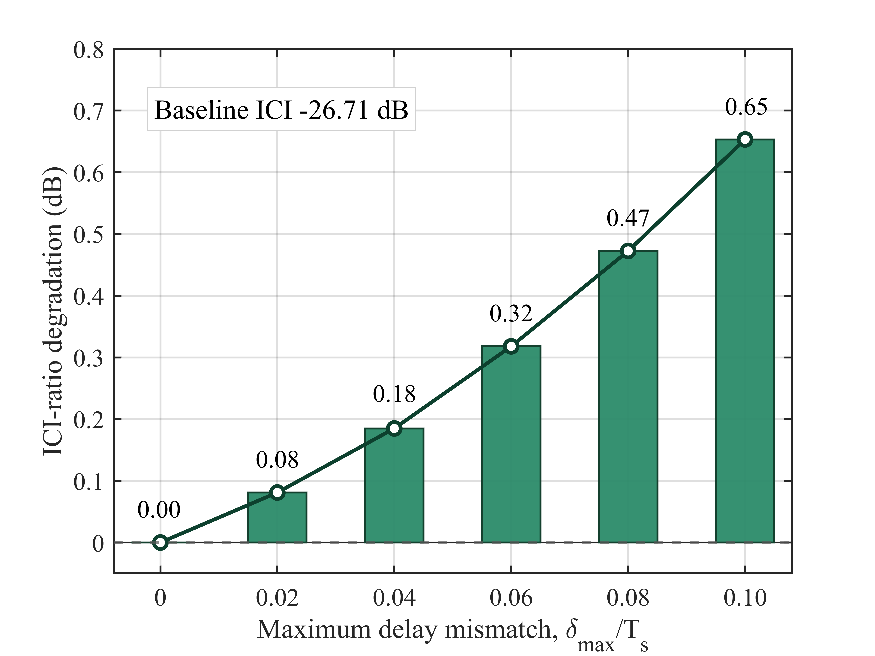}}
    \vspace{-0.2 cm}
\caption{Robustness of the proposed wave-domain linear receiver. 
(a) Effective spectral efficiency versus residual timing offset. 
(b) Timing-induced OFDM leakage measured by the ICI ratio, where lower values indicate better subcarrier decoupling. 
(c) Effective adjacent-delay perturbation under the paired sweep in \eqref{eq:delay_profile_perturbation}, measured by the ICI-ratio degradation relative to the nominal channel profile.}
    \label{fig:robustness}
\vspace{-0.7 cm}
\end{figure*}

We consider two representative stress tests: residual timing offset after synchronization and effective adjacent-delay perturbation of the channel profile. For timing robustness, the single-hypothesis SIM$_2$ uses only the nominal timing window, whereas the multi-hypothesis SIM$_2$ follows \eqref{eq:multi_hypothesis_oa} and \eqref{eq:timing_hypothesis_selection}. Fig.~\ref{fig:misa} shows that the multi-hypothesis mode maintains high spectral efficiency over \(\Delta\tau\in\{-1,0,1\}\), where the conventional digital DFT+MMSE receiver and the single-hypothesis SIM$_2$ are more sensitive to window shift. This result supports the synchronization interface in Section~\ref{sec:e2e_receiver_matrix}: the control branch only needs to provide a coarse timing reference, while SIM$_2$ can expose several neighboring cyclic windows and select the one with the smallest pilot residual. Fig.~\ref{fig:misb} shows the same trend in ICI ratio, confirming that the improvement comes from better subcarrier decoupling rather than detector advantage. In other words, residual timing robustness is obtained at the OFDM-analysis stage, before SIM$_3$ performs residual MIMO detection. For delay-profile robustness, we use the adjacent-tap leakage model in \eqref{eq:delay_profile_perturbation}. A paired sweep fixes the channel and perturbation directions while increasing the normalized perturbation level \(\epsilon_{\tau}\) from \(0\) to \(0.1\). As shown in Fig.~\ref{fig:misc}, the ICI-ratio degradation remains below \(1\) dB even at \(\epsilon_{\tau}=0.1\). This indicates that the proposed receiver does not rely on an exactly matched delay profile to maintain OFDM-domain decoupling. Therefore, the perturbation result should be interpreted as a communication-layer robustness test for channel-delay uncertainty, rather than as a detailed circuit-level DAM mismatch model.

\vspace{-0.3 cm}
\section{Conclusion}
\label{sec:conclusion}

This paper proposed a SIM-enabled wave-domain linear receiver for MIMO-OFDM systems, showing that wave-domain architecture can serve as a physical computing substrate for linear receiver design. The proposed architecture redistributes the main linear modules of a conventional receiver, including pre-DFT channel-memory processing, CP-windowed DFT analysis, and residual MIMO linear detection, into the SIM-enabled receive path. SIM$_1$ realizes channel-adaptive temporal memory processing, SIM$_2$ materializes OFDM analysis through useful-window memory and Fourier-coded wave-domain summation, and SIM$_3$ performs block-structured virtual-subcarrier detection. Numerical results validated the functional roles of the three wave-domain operators, their end-to-end performance under CP-insufficient and CP-sufficient regimes, and their robustness to residual timing offsets and effective channel-delay-profile perturbations. These results support the feasibility of receiver-side wave-domain OFDM analysis and motivate further studies on hardware-calibrated prototypes, coded-link evaluation, and joint wave-domain transceiver co-design.


\vspace{-0.5cm}
\begin{thebibliography}{99}


\bibitem{OFDM}
H. Bolcskei, ``MIMO-OFDM wireless systems: Basics, perspectives, and challenges,'' \emph{IEEE Wireless Commun.}, vol. 13, no. 4, pp. 31--37, Aug. 2006.

\bibitem{OFDM2}
G. L. Stuber, J. R. Barry, S. W. McLaughlin, Y. Li, M. A. Ingram, and T. G. Pratt, ``Broadband MIMO-OFDM wireless communications,'' \emph{Proc. IEEE}, vol. 92, no. 2, pp. 271--294, Feb. 2004.

\bibitem{vision1}
C. Wang, X. You, X. Gao \emph{et al.}, ``On the road to 6G: Visions, requirements, key technologies, and testbeds,'' \emph{IEEE Commun. Surveys Tuts.}, vol. 25, no. 2, pp. 905--974, 2023.

\bibitem{vision2}
H. Tataria, M. Shafi, A. F. Molisch, M. Dohler, H. Sjoland, and F. Tufvesson, ``6G wireless systems: Vision, requirements, challenges, insights, and opportunities,'' \emph{Proc. IEEE}, vol. 109, no. 7, pp. 1166--1199, July 2021.

\bibitem{HY1}
A. F. Molisch, V. V. Ratnam, S. Han \emph{et al.}, ``Hybrid beamforming for massive MIMO: A survey,'' \emph{IEEE Commun. Mag.}, vol. 55, no. 9, pp. 134--141, Sep. 2017.

\bibitem{HY2}
S. Park, A. Alkhateeb, and R. W. Heath, ``Dynamic subarrays for hybrid precoding in wideband mmWave MIMO systems,'' \emph{IEEE Trans. Wireless Commun.}, vol. 16, no. 5, pp. 2907--2920, May 2017.

\bibitem{HY3}
F. Sohrabi and W. Yu, ``Hybrid analog and digital beamforming for mmWave OFDM large-scale antenna arrays,'' \emph{IEEE J. Sel. Areas Commun.}, vol. 35, no. 7, pp. 1432--1443, Jul. 2017.

\bibitem{SIM0}
J. An \emph{et al.}, ``Stacked intelligent metasurface-aided MIMO transceiver design,'' \emph{IEEE Wireless Commun.}, vol. 31, no. 4, pp. 123--131, Aug. 2024.

\bibitem{Lin2018D2NN}
X. Lin, Y. Rivenson, N. T. Yardimci, M. Veli, Y. Luo, M. Jarrahi, and A. Ozcan, ``All-optical machine learning using diffractive deep neural networks,'' \emph{Science}, vol. 361, no. 6406, pp. 1004--1008, Jul. 2018.

\bibitem{Liu2022D2NN}
C. Liu \emph{et al.}, ``A programmable diffractive deep neural network based on a digital-coding metasurface array,'' \emph{Nature Electronics}, vol. 5, no. 2, pp. 113--122, Feb. 2022.

\bibitem{An2023SIM}
J. An, C. Yuen, L. Dai, M. Di Renzo, M. Debbah, and L. Hanzo, ``Stacked intelligent metasurfaces for efficient holographic MIMO communications in 6G,'' \emph{IEEE J. Sel. Areas Commun.}, vol. 41, no. 8, pp. 2380--2396, Aug. 2023.

\bibitem{Papazafeiropoulos_TWC_2024_HolographicSIM}
A. Papazafeiropoulos, J. An, P. Kourtessis, T. Ratnarajah, and S. Chatzinotas, ``Achievable rate optimization for stacked intelligent metasurface-assisted holographic MIMO communications,'' \emph{IEEE Trans. Wireless Commun.}, vol. 23, no. 10, pp. 13173--13186, Oct. 2024.

\bibitem{Li1}
Z. Li, J. An, and C. Yuen, ``Stacked intelligent metasurfaces for fully-analog wideband beamforming design,'' in \emph{Proc. IEEE APWCS}, Singapore, Aug. 2024, pp. 1--5.

\bibitem{Hassan_OJCOMS_2024_Beamforming}
N. U. Hassan, J. An, M. Di Renzo, M. Debbah, and C. Yuen, ``Efficient beamforming and radiation pattern control using stacked intelligent metasurfaces,'' \emph{IEEE Open J. Commun. Soc.}, vol. 5, pp. 599--611, Jan. 2024.

\bibitem{An_TWC_2025_MUBeamforming}
J. An \emph{et al.}, ``Stacked intelligent metasurfaces for multiuser downlink beamforming in the wave domain,'' \emph{IEEE Trans. Wireless Commun.}, vol. 24, no. 7, pp. 5525--5538, Jul. 2025.

\bibitem{Nadeem_WCNC_2024_ChannelEst}
A. Lawal, A. Zerguine, A. A. Nasir, and K. Abed-Meraim, ``Channel estimation for stacked intelligent metasurface aided network using deep learning,'' \emph{IEEE Commun. Lett.}, vol. 29, no. 11, pp. 2641--2645, Nov. 2025.

\bibitem{Yao_WCL_2024_ChannelEst}
X. Yao, J. An, L. Gan, M. Di Renzo, and C. Yuen, ``Channel estimation for stacked intelligent metasurface-assisted wireless networks,'' \emph{IEEE Wireless Commun. Lett.}, vol. 13, no. 5, pp. 1349--1353, May 2024.

\bibitem{Papazafeiropoulos_WCL_2024_NearField}
A. Papazafeiropoulos, P. Kourtessis, S. Chatzinotas, D. I. Kaklamani, and I. S. Venieris, ``Near-field beamforming for stacked intelligent metasurfaces-assisted MIMO networks,'' \emph{IEEE Wireless Commun. Lett.}, vol. 13, no. 11, pp. 3035--3039, Nov. 2024.

\bibitem{Niu_TIFS_2024_SecureSIM}
H. Niu, X. Lei, J. An, L. Zhang, and C. Yuen, ``On the efficient design of stacked intelligent metasurfaces for secure SISO transmission,'' \emph{IEEE Trans. Inf. Forensics Security}, vol. 20, pp. 60--70, Nov. 2024.

\bibitem{Niu_WCL_2024_ISACSIM}
H. Liu \emph{et al.}, ``Stacked intelligent metasurfaces for wireless sensing and communication: Applications and challenges,'' \emph{IEEE Wireless Commun.}, vol. 32, no. 4, pp. 46--53, Aug. 2025.

\bibitem{Huang_WCL_2025_SemanticSIM}
G. Huang, J. An, Z. Yang, L. Gan, M. Bennis, and M. Debbah, ``Stacked intelligent metasurfaces for task-oriented semantic communications,'' \emph{IEEE Wireless Commun. Lett.}, vol. 14, no. 2, pp. 310--314, Feb. 2025.

\bibitem{Li_TCOM_2024_CellFreeSIM}
Q. Li \emph{et al.}, ``Stacked intelligent metasurfaces for holographic MIMO aided cell-free networks,'' \emph{IEEE Trans. Commun.}, vol. 72, no. 11, pp. 7139--7151, Nov. 2024.

\bibitem{survey1}
M. Liu, C. Yuen, D. Niyato, B. Clerckx, and L. Hanzo, ``In-wave computation aided stacked intelligent metasurfaces in next-generation networks: Challenges and opportunities,'' \emph{IEEE Wireless Commun.}, 2026, doi: 10.1109/MWC.2026.3672362.

\bibitem{survey2}
H. Niu, C. Yuen, M. Di Renzo, M. Debbah, and H. V. Poor, ``Two-layer stacked intelligent metasurfaces: Balancing performance and complexity,'' \emph{IEEE Wireless Commun.}, 2026, doi: 10.1109/MWC.2026.3674018.

\bibitem{NN}
K. Stylianopoulos, M. E. Pandolfo, P. Di Lorenzo, and G. C. Alexandropoulos, ``Metasurfaces-integrated wireless neural networks for lightweight over-the-air edge inference,'' \emph{IEEE Wireless Commun.}, 2026, doi: 10.1109/MWC.2026.3682709.

\bibitem{survey3}
J. Zhang, Q. Li \emph{et al.}, ``Spherical stacked intelligent metasurfaces: A paradigm for full-space wave-domain processing,'' \emph{IEEE Wireless Commun.}, 2026, doi: 10.1109/MCOM.001.2600020.

\bibitem{Li2}
Z. Li, J. An, and C. Yuen, ``Stacked intelligent metasurfaces-enhanced MIMO OFDM wideband communication systems,'' \emph{IEEE Trans. Wireless Commun.}, vol. 25, pp. 9608--9622, Jan. 2026.

\bibitem{Li3}
Z. Li, J. An, and C. Yuen, ``Fundamental trade-off in wideband stacked intelligent metasurface assisted OFDMA systems,'' in \emph{Proc. IEEE GLOBECOM}, Taipei, Taiwan, Dec. 2025, pp. 2384--2389.

\bibitem{Li4}
Z. Li, J. An, and C. Yuen, ``Stacked intelligent metasurface-enhanced wideband multiuser MIMO OFDM-IM communications,'' \emph{IEEE Trans. Wireless Commun.}, 2026, doi: 10.1109/TWC.2026.3713490.

\bibitem{QLi}
Q. Li, M. El-Hajjar, C. Xu, J. An, C. Yuen, and L. Hanzo, "Stacked intelligent metasurface-based transceiver design for near-field wideband systems," \emph{IEEE Trans. Wireless Commun.}, vol. 73, no. 9, pp. 8125-8139, Sept. 2025

\bibitem{Li5}
Z. Li, J. An, and C. Yuen, ``Baseband-free wideband MU-MIMO OFDM via stacked intelligent metasurfaces,'' accepted in \emph{IEEE GLOBECOM 2026}, Available: https://arxiv.org/abs/2607.21900.

\bibitem{Li6}
Z. Li, J. An, and C. Yuen, ``A fully wave-domain wideband MU MIMO OFDM transmitter via stacked intelligent metasurfaces,'' Available: https://arxiv.org/abs/2609.09727.

\bibitem{DMA}
H. Wang \emph{et al.}, ``Dynamic metasurface antennas for MIMO-OFDM receivers with bit-limited ADCs," \emph{IEEE Trans. Commun.}, vol. 69, no. 4, pp. 2643--2659, Apr. 2021.

\bibitem{NBF}
X. Teng, W. Tang, X. Li, and S. Jin, ``Baseband-free end-to-end communication system based on diffractive deep neural network,'' \emph{IEEE Trans. Wireless Commun.}, vol. 25, pp. 7649--7664, Jan. 2026.

\bibitem{DRL}
H. Liu, J. An, G. C. Alexandropoulos, D. W. K. Ng, C. Yuen, and L. Gan, ``Multi-user MISO with stacked intelligent metasurfaces: A DRL-based sum-rate optimization approach,'' \emph{IEEE Trans. Cognitive Commun. Netw.}, vol. 12, pp. 251--266, 2026.

\bibitem{DRL2} 
E. Shi \emph{et al.}, ``Low-complexity multi-agent continual learning for stacked intelligent metasurface-assisted secure communications,'' \emph{arXiv:2602.01653}, Feb. 2026.

\bibitem{Nakanishi2013EITStorage}
T. Nakanishi, T. Otani, Y. Tamayama, and M. Kitano, ``Storage of electromagnetic waves in a metamaterial that mimics electromagnetically induced transparency,'' \emph{Phys. Rev. B}, vol. 87, no. 16, p. 161110, Apr. 2013.

\bibitem{Nakanishi2018EITRetrieval}
T. Nakanishi and M. Kitano, ``Storage and retrieval of electromagnetic waves using electromagnetically induced transparency in a nonlinear metamaterial,'' \emph{Appl. Phys. Lett.}, vol. 112, no. 20, p. 201905, May 2018.

\bibitem{An2024AdjustableDelayRIS}
J. An, C. Xu, D. W. K. Ng, C. Yuen, and L. Hanzo, ``Adjustable-delay RIS is capable of improving OFDM systems,'' \emph{IEEE Trans. Veh. Technol.}, vol. 73, no. 7, pp. 9927--9942, Jul. 2024.

\end{thebibliography}
\end{document}